\documentclass{article}
\usepackage[a4paper, margin=0.7in]{geometry}
\usepackage{natbib}
\setcitestyle{authoryear,open={(},close={)}} 
\usepackage[colorlinks=true,urlcolor=blue,citecolor=red,linkcolor=red,bookmarks=true]{hyperref}
\usepackage{graphicx}
\graphicspath{ {./images/} }

\usepackage{hyperref}
\usepackage{multirow}

\usepackage{algorithm}
\usepackage{algpseudocode}

\usepackage{amssymb,amsmath}

\usepackage{enumitem}

\usepackage{xcolor}
\def\HiLi{\leavevmode\rlap{\hbox to \hsize{\color{yellow!50}\leaders\hrule height .8\baselineskip depth .5ex\hfill}}}

\title{Precision and Decisiveness as Goals: Reliable Sequential Hypothesis Testing with a Dual Stopping Criterion}
\date{August 04, 2026}
\author{Eyal A. Kazin}

\begin{document}
\maketitle

Sequential hypothesis testing offers flexibility over fixed-sample designs, but stopping rules coupled to decision criteria risk confirmation bias through early peeking.
The HDI+ROPE algorithm exemplifies this: it stops as soon as the posterior Highest Density Interval (HDI) falls entirely outside a Region of Practical Equivalence (ROPE) around the null (rejecting it) or entirely inside, which enables positive acceptance of the null, something Null Hypothesis Significance Testing cannot do. However, this coupling means a wide HDI can satisfy the criterion on early, unrepresentative samples, incurring a systematic false-positive cost.
Fully decoupled methods, such as ``Precision is the Goal'' (PitG),
eliminate this bias by halting only once a target HDI width $\omega_{\rm goal}$ is reached;
yet, by divorcing the stopping rule from the decision criterion, PitG frequently yields inconclusive outcomes, particularly when the null hypothesis is true.
We propose ``Decisive Precision is the Goal'' (DPitG), which requires both the precision target and a conclusive verdict to be satisfied simultaneously.
In fair coin simulations ($\omega_{\rm goal}=0.08$, ROPE$=0.5\pm0.05$), DPitG reduces the PitG inconclusive rate from 62\% to 2\% at a median cost of only 5\% more samples, with zero false positives;
HDI+ROPE achieves comparable conclusiveness only at the cost of false positives.
Across the $\omega_{\rm goal}$ range tested, DPitG conclusiveness remains above 97\% while PitG's varies widely.
We provide a closed-form planning formula
$N_{\rm goal} \propto V/\omega_{\rm goal}^2$ (where $V$ is the observation variance),
an online interactive calculator, and open-source code.
Demonstrated on single-group binary data, the framework extends readily to continuous outcomes and two-group comparisons.
DPitG's fully pre-specified stopping rule is compatible with pre-registration standards
and is the method of choice whenever a reliable verdict is required
provided the study can afford to collect at least $N_{\rm goal}$ observations.

\section{Introduction}

Empirical hypothesis testing is foundational to scientific progress, enabling researchers
to assess how well a sample represents an underlying truth.
A central challenge in this process is determining when enough data have been collected
to make a reliable decision, especially as the accuracy of statistical assessments
depends heavily on sample size.  Careful planning for statistical power and precision
is essential to avoid biased inference and ensure meaningful results \citep{cohen1988}.

Sequential hypothesis testing offers a flexible alternative to fixed-sample designs
\citep{wald1947} by allowing data to be evaluated as it accumulates.
This approach is widely used in fields such as clinical trials
\citep{jennison2000, fda2010bayesian}, quality control, A/B testing \citep{johari2022}, and financial
monitoring,
offering potential benefits in efficiency, timeliness, and ethical considerations.
Central to this approach is the distinction between two critical components.
The \textit{stopping rule} dictates when data collection ends; the
\textit{decision rule} then determines the final verdict on the hypothesis,
one of three outcomes: accept, reject, or inconclusive.
While these components are conceptually distinct, many operational methods
conflate them, triggering a stop exactly when a decision criterion is met.

This simultaneous application of stopping and decision rules is common in widely used
frequentist and Bayesian methods, such as p-value thresholds \citep{fisher1925} or posterior-based heuristics
(e.g., HDI+ROPE \citealp{kruschke2011, kruschke2013, kruschke2015doing},
Bayes Factors \citealp{kassraftery1995}).
However, if not carefully designed, this coupling can lead to confirmation bias:
premature stopping on extreme or unrepresentative early samples, a phenomenon known
as early peeking \citep{simmons2011}.
As a concrete illustration: in the fair coin setting
($\theta_{\rm true}=\theta_{\rm null}=0.5$, credible interval of 95\% and minimum effect size $\pm 0.05$),
a sequentially monitored HDI+ROPE algorithm achieves high conclusiveness
but at a systematic 6.3\% false-positive rate (Section~\ref{sec:fair_coin}).

One principled remedy is to decouple the stopping criterion entirely from the hypothesis
outcome, halting based on posterior precision rather than the direction of the result.

Following the \textit{Accuracy in Parameter Estimation} paradigm \citep{maxwell2008},
\cite{kruschke2015doing} advocated for a method that strictly decouples these two steps,
thereby addressing the risk of premature and biased conclusions from early peeking.
He proposed using a predetermined target for posterior precision as the
sole \textit{stopping rule}, independent of the hypothesis outcome. Only once this precision goal
is met is the \textit{decision rule} applied to accept or reject the null hypothesis
(via a posterior credible interval criterion).
\cite{kruschke2015doing} demonstrated that this ``Precision is the Goal'' (PitG) method
eliminates the early-peeking bias in outcome sampling. However, our analysis reveals an important
limitation: when the null hypothesis is true, or practically so, PitG often yields a high rate of inconclusive
results, rather than correctly accepting the null.\footnote{\citealt{kruschke2015doing} noted in passing that PitG ``did remain undecided
in some cases'' (\S13.3.2, page 391); our analysis shows this is a systematic and
striking effect, with the sampling budget $N_{\rm max}$ and precision goal $\omega_{\rm goal}$
as the key governing parameters. See below for quantitative examples.}

In this paper, we propose an improved variant we call
``Decisive Precision is the Goal'' (DPitG), in which precision is a necessary but no
longer sufficient stopping condition: data collection continues until \textit{both} the
precision goal and a decisive hypothesis outcome are satisfied simultaneously.
The benefit is most pronounced precisely where PitG's inconclusive rate is highest,
namely when the null is true or practically so, making DPitG particularly valuable for
confirming the null hypothesis.
While scientific incentives often favour the discovery of new effects,
rigorous confirmation of the null is equally critical in many fields: from bioequivalence
trials for generic drugs \citep{schuirmann1987, karalis2012}
to do-no-harm regression testing in software
deployment and futility stopping in programme evaluations.
Unlike \textit{Null Hypothesis Significance Testing} (NHST, see Appendix~\ref{app:nhst}),
which can only fail to reject the null, the precision-based framework enables positive
acceptance as a genuine and actionable verdict, and DPitG ensures this verdict is
reliably reached.
This fully pre-specified rule is also compatible with pre-registration standards
\citep{nosek2018} and directly addresses the design-stage imprecision implicated in
the replication crisis \citep{osc2015, gelman2014}.

We compare DPitG and PitG against the HDI+ROPE algorithm, a coupled baseline that stops as
soon as the posterior yields a decisive result, through simulations with
single-group dichotomous data.\footnote{Two other widely used approaches, NHST
(p-value-based) and Bayes Factors, are not among the three algorithms studied
here. Both lack an explicit effect-size criterion (the ROPE); NHST cannot formally
accept the null, while Bayes Factors lack a direct precision goal.
Their limitations are illustrated in Appendices~\ref{app:nhst}
and~\ref{app:bayes_factors}, respectively.}
Applied to this setting, DPitG substantially outperforms both alternatives: for
example, at $\omega_{\rm goal}=0.08$, it reduces PitG's $62.1\%$ inconclusive rate
to $2.2\%$ at a median cost of only $4.7\%$ extra samples, while preserving a zero
false-positive rate, an advantage the HDI+ROPE algorithm matches only at a $6.2\%$
false-positive cost (Section~\ref{sec:fair_coin}). These advantages hold broadly
across precision goals and true parameter values
(Sections~\ref{sec:fair_coin}--\ref{sec:trends}).
The framework extends to continuous outcomes
(Appendix~\ref{app:expected_stop}) and two-group comparisons
(Appendix~\ref{app:extensions}).
Alongside the empirical results we provide a closed-form planning formula, an online
interactive calculator, and open-source code to support adoption in applied settings.

Section~\ref{sec:methods} describes the theoretical framework and the three stopping
algorithms; Section~\ref{sec:results} presents simulation results; and
Section~\ref{sec:discussion} places the findings in methodological context.

\section{Methods}\label{sec:methods}
We study three algorithms built on two summary properties of the posterior
distribution: its \textit{location} (how far the estimate sits from the null,
conveying the effect size) and its \textit{width} (how uncertain that estimate
is)\footnote{Here \textit{epistemic uncertainty} means the uncertainty in our
knowledge about the true value of a parameter, as opposed to
\textit{aleatory uncertainty}, which is the inherent variability of the system
being studied.}.
Location is formalised by where the posterior credible interval sits relative to a
pre-specified equivalence region around the null; width by how broad that interval is.
Together they inform a shared Decision Rule that all three algorithms apply once
their respective stopping criteria are met.

Section~\ref{sec:shared_components} introduces the shared inference components,
namely the ROPE, the HDI, and the Decision Rule, that all three algorithms apply
identically.
Section~\ref{sec:sequential_stopping_algorithms} describes the three algorithms,
which differ only in their stopping criteria.
Section~\ref{sec:expected_stop_iteration} derives the expected stopping iteration for
precision-based methods analytically.
Section~\ref{sec:experimental_design} details the simulation design used to evaluate
and compare all three algorithms.

\subsection{Shared Components}\label{sec:shared_components}

\subsubsection{Effect Size by Region of Practical Equivalence (ROPE)}\label{sec:rope}

A common misconception in hypothesis testing is that a ``statistically significant'' outcome is sufficient for real-world decision-making. In practice, one must consider the \textit{effect size} \citep{cohen1988}: how large is the difference, not merely whether it is detectable. For example, medical research requires a \textit{minimal clinically important difference} (MCID), the smallest change that a patient or clinician would consider meaningful, for an outcome to justify a change in treatment \citep{jaeschke1989}.

Consider a hypothetical scenario where a researcher examines whether a therapeutic
differs in impact by gender.
Even if the data yield a ``statistically significant'' result (say, the drug is
72.1\% beneficial for males and 72.3\% for females), a practitioner may consider
this equivalent for all practical purposes.
Hence, the researcher should define in advance a pre-specified equivalence margin
around the null hypothesis.

One principled way to operationalise effect size is the
\textit{Region of Practical Equivalence} (ROPE; \citealt{kruschke2011}).
The ROPE is defined as an area around the null value considered ``similar enough to the null hypothesis''
such that if the true value is within this area, it is effectively the same as the null hypothesis.

The appropriate ROPE width is domain-specific: a board game manufacturer requires
dice to be fair to a reasonable standard, but not to a precision indistinguishable
to a casual player; conversely, a casino requires a much stricter tolerance (a narrower
ROPE) to ensure regulatory compliance and fair play.
Appendix~\ref{app:practical_guidance} discusses principled approaches for setting
the ROPE, including domain standards such as the MCID and bioequivalence guidelines,
and robustness checks across a range of ROPE widths when no domain standard exists.

\subsubsection{Highest Density Interval (HDI)}\label{sec:precision_hdi}

To quantify the uncertainty in our estimate we use the \textit{credible interval},
the Bayesian analogue of the frequentist confidence interval:
the region of the parameter space that contains a specified fraction of the posterior mass
(e.g.\ 95\%).
Several rules exist for selecting which region to report; the two most common are the
\textit{Equal-Tailed Interval} (ETI) and the \textit{Highest Density Interval} (HDI)
(see Appendix~\ref{app:hdi} for their formal definitions).

The HDI has a natural geometric interpretation: imagine drawing a horizontal line across
the posterior density curve and slowly lowering it from the peak. The line intersects the
curve at two points, enclosing an area underneath. The HDI is the interval between those
two intersection points at the moment the enclosed area first reaches the target mass
(e.g.\ 95\%). Because the line is horizontal, every point inside the interval sits on a
portion of the curve that is at least as tall as any point outside — hence \textit{highest
density}. For a symmetric, bell-shaped posterior this coincides with the ETI (equal tails),
but for a skewed posterior the HDI is noticeably shorter.

Throughout this work we use the HDI at the 95\% level to remain comparable with traditional
frequentist hypothesis testing.\footnote{\cite{kruschke2015doing} notes that for numerical posteriors
obtained via Markov Chain Monte Carlo (MCMC), 95\% HDI estimation requires at least 10{,}000 samples; a slightly lower
level such as 94\% is recommended for fewer samples. \cite{mcelreath2016} advocates 89\%
(the highest prime below 100, to underscore the arbitrariness of any threshold). Because we employ analytical
posteriors in our experiments (Section~\ref{sec:experimental_design}), sampling noise is not a
concern and 95\% is unambiguous; practitioners relying on sampled (e.g., MCMC-based) posteriors
should consider the lower thresholds noted above.}

\subsubsection{The Decision Rule}\label{sec:decision_criterion}

We distinguish between the \textit{Stopping Rule} (when to stop collecting data)
and the \textit{Decision Rule} (what to conclude once stopped). While the three algorithms
discussed later differ in their \textit{stopping} criteria, they all share an identical
Decision Rule.

The decision is determined solely by the
relationship between the HDI and the ROPE.
Let ${\rm ROPE} = [{\rm ROPE}_{\rm min}, {\rm ROPE}_{\rm max}]$ and the posterior HDI limits be
$[{\rm HDI}_{\rm min}, {\rm HDI}_{\rm max}]$,
we define three outcomes based on their relative positions:

\begin{enumerate}[label=(\alph*)]
    \item \textit{Accept Null}: The HDI is completely \textit{inside} the ROPE. \\
    $({\rm ROPE}_{\rm min} \leq {\rm HDI}_{\rm min}) \land ({\rm HDI}_{\rm max} \leq {\rm ROPE}_{\rm max})$
    \item \textit{Reject Null}: The HDI is completely \textit{outside} the ROPE. \\
    $ ({\rm HDI}_{\rm max} < {\rm ROPE}_{\rm min}) \lor ({\rm ROPE}_{\rm max} < {\rm HDI}_{\rm min} )$
    \item \textit{Inconclusive}: The HDI \textit{straddles} the ROPE boundary. \\ Otherwise.
\end{enumerate}

The interpretation of these outcomes, particularly what it means to \textit{accept}
the null in a Bayesian framework, differs meaningfully from the frequentist convention
and is worth examining carefully; we do so in Section~\ref{sec:accept_reject_meaning}.

\begin{figure}[h!]
    \centering
    \includegraphics[width=1\textwidth]{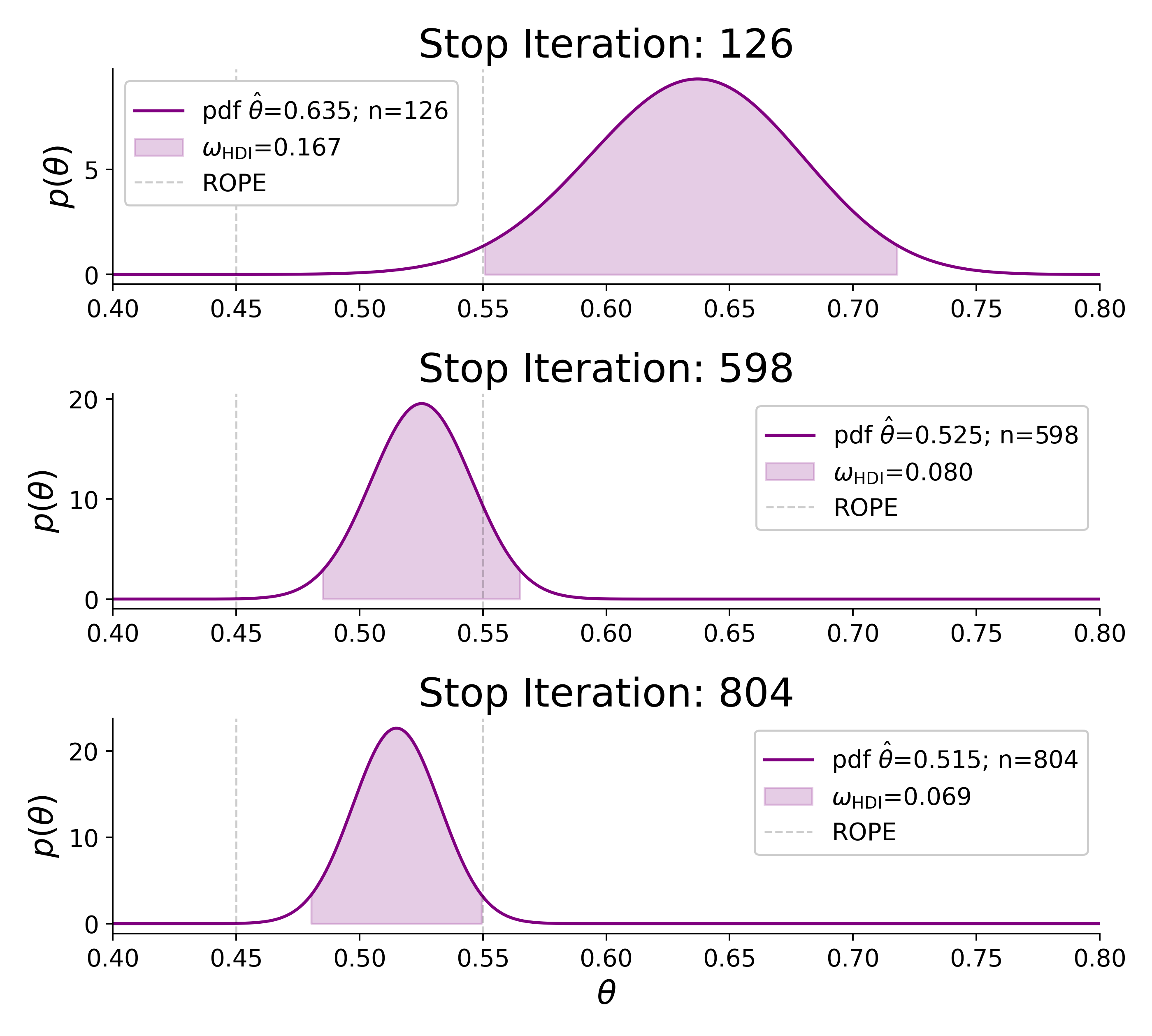}
    \caption{Posteriors at three stop iterations from a hand-picked experiment
    illustrating the three possible decision outcomes.
    Shaded areas are 95\% HDIs; their widths $\omega_{\rm HDI}$ are annotated.
    Vertical dashed lines mark the ROPE boundaries.
    \textit{Top} (iteration 126): HDI fully outside the ROPE $\rightarrow$ Reject $\theta_{\rm null}$.
    \textit{Middle} (iteration 598): HDI straddles the ROPE $\rightarrow$ Inconclusive.
    \textit{Bottom} (iteration 804): HDI fully within the ROPE $\rightarrow$ Accept $\theta_{\rm null}$.
    See Figure~\ref{fig:iterations} for the full sequence of iterations for this experiment.
    }
    \label{fig:posteriors}
\end{figure}

Figure~\ref{fig:posteriors} illustrates all three outcomes using a hand-picked Bernoulli
experiment ($\theta_{\rm null}=0.5$, ${\rm ROPE}_{\rm min}^{\rm max}=[0.45,0.55]$).
In the top panel, a stop at iteration 126 finds the 95\% HDI fully outside the ROPE, leading to rejection of the null hypothesis.
In the middle panel, a stop at iteration 598 finds the HDI straddling the ROPE boundary, resulting in an inconclusive decision.
In the bottom panel, a stop at iteration 804 finds the 95\% HDI fully within the ROPE, allowing us to accept $\theta_{\rm null}$.

Figure~\ref{fig:posteriors} illustrates that the decision outcome may be iteration-dependent. An inconclusive outcome implies that the data collected is insufficient to justify a decision. In a live setting, a human-in-the-loop (or an algorithm, common in automated high-frequency environments) would need to either conduct a risk assessment or collect more data. Since risk assessment is highly contextual, we defer that discussion and focus on the necessity of collecting more data to avoid inconclusive outcomes.

Algorithm \ref{alg:decision_criterion} further formalises this logic.

\begin{algorithm}
    \caption{The Decision Rule}\label{alg:decision_criterion}
    \begin{algorithmic}
    \Require $\mathrm{ROPE}_\mathrm{min}$, $\mathrm{ROPE}_\mathrm{max}$, $\mathrm{HDI}_\mathrm{min}$, $\mathrm{HDI}_\mathrm{max}$
    \If{$(\mathrm{ROPE}_\mathrm{min} \leq \mathrm{HDI}_\mathrm{min}) \ \& \ (\mathrm{HDI}_\mathrm{max} \leq \mathrm{ROPE}_\mathrm{max})$}
        \State Decision = Accept $\theta_{\rm null}$ \Comment{HDI completely within ROPE}
    \ElsIf{$\mathrm{ROPE}_\mathrm{max}<\mathrm{HDI}_\mathrm{min}$}
        \State Decision = Reject $\theta_{\rm null}$ \Comment{HDI completely outside ROPE; $\theta_{\rm null}<\hat\theta$}
    \ElsIf{$\mathrm{HDI}_\mathrm{max}< \mathrm{ROPE}_\mathrm{min}$}
        \State Decision = Reject $\theta_{\rm null}$ \Comment{HDI completely outside ROPE; $\hat\theta<\theta_{\rm null}$}
    \Else
    \State Decision = Inconclusive  \Comment{HDI straddles ROPE}
    \EndIf \\
    \Return Decision
    \end{algorithmic} 
\end{algorithm}

\subsubsection{Interpreting Accept and Reject Decisions}\label{sec:accept_reject_meaning}

A key conceptual distinction between Bayesian and frequentist inference lies in what it
means to \textit{accept} the null hypothesis. In NHST, a small p-value justifies
\textit{rejecting} the null; a large p-value only means there is insufficient evidence
to reject it. One cannot formally \textit{accept} the null, only ``fail to reject'' it.
Absence of evidence is not evidence of absence: NHST can conclusively reject
the null but cannot deliver its symmetric counterpart, a positive verdict that
the null actually holds.

The HDI+ROPE framework resolves this asymmetry: because the ROPE encodes a pre-specified
notion of practical equivalence, each verdict carries positive evidential weight:
Accept and Reject are both affirmative findings, not merely the presence or absence of a
signal.\footnote{Note that Bayes Factors can also yield positive evidence for the null, but do so
by comparing marginal likelihoods rather than by assessing practical equivalence
(see Appendix~\ref{app:bayes_factors}).} The HDI+ROPE framework's Accept verdict is therefore
distinctive in being grounded in a pre-specified effect-size criterion.

Neither Accept nor Reject identifies the sample estimate $\hat\theta$ as the
true value; both are statements about $\theta_{\rm null}$, not $\hat\theta$. The posterior
distributes probability continuously across parameter space, and the HDI+ROPE framework distils
this into a verdict by asking whether the bulk of posterior mass lies inside, outside, or
straddling the ROPE. The full posterior remains available for a more nuanced reading: in
particular, a ``Reject null'' decision does not endorse $\hat\theta$ as the true value;
$\hat\theta$ is the mode (or mean) of the posterior, and the full distribution
quantifies the uncertainty surrounding it.

The panels of Figure~\ref{fig:posteriors} make this concrete under the default setup
($\theta_{\rm null}=0.5$, $\mathrm{ROPE}_{\rm max}^{\rm min}=[0.45,\,0.55]$):

\begin{enumerate}[label=(\alph*)]
  \item Bottom panel (iteration 804, Accept):
        $\hat\theta=0.515$ and the 95\% HDI is
        $[0.480,\,0.549]$.
        Both bounds lie inside the ROPE, so we accept
        $\theta_{\rm null}=0.5$: positive evidence that the true rate is practically
        equivalent to $0.5$. We are \textit{not} claiming the true rate equals $0.515$.
  \item Top panel (iteration 126, Reject):
        $\hat\theta=0.635$ and the 95\% HDI is
        $[0.551,\,0.718]$.
        Both bounds lie above $\mathrm{ROPE}_{\rm max}=0.55$, so we reject
        $\theta_{\rm null}=0.5$: positive evidence that the true rate is meaningfully
        above $0.5$. We are \textit{not} claiming the true rate equals $0.635$.
\end{enumerate}

Two error types follow directly.
A \textit{false-accept} arises when a conclusive Accept verdict is reached but
$\theta_{\rm true}$ lies outside the ROPE, the HDI+ROPE framework analogue of a false
negative, since the data appear to confirm equivalence when it does not hold.
A \textit{false-reject} arises when a conclusive Reject verdict is reached but
$\theta_{\rm true}$ lies inside the ROPE, the analogue of a false positive, since
the data appear to refute a null that is in fact practically true.
Their rates across $\theta_{\rm true}$ values are characterised in
Section~\ref{sec:results}.

\subsection{Sequential Stopping Algorithms}\label{sec:sequential_stopping_algorithms}

Table \ref{tab:comparison} provides a high-level comparison of the three algorithms.
Each algorithm shares the same Decision Rule (Algorithm \ref{alg:decision_criterion})
but differs in when it triggers the stop condition.
Because the term ``HDI+ROPE'' serves two distinct roles in this work, we use distinct names
throughout. \textit{HDI+ROPE framework} denotes the shared Decision Rule applied
identically by all three algorithms (Section~\ref{sec:decision_criterion}).
\textit{HDI+ROPE algorithm} (equivalently, \textit{HDI+ROPE stopping rule}) denotes
the specific sequential stopping algorithm described in
Section~\ref{sec:hdi_rope_stopping}.

\begin{table}[h!]
    \begin{center}
      \begin{tabular}{l|c|l|l}
        \textbf{Algorithm} & \textbf{Key Input} & \textbf{Stop Criterion} &  \textbf{Precision/Location Coupling} \\
        \hline
        HDI+ROPE & $N_\mathrm{min}$ & Decision is conclusive & Location only \\
        Precision is the Goal & $\omega_{\rm Goal}$ & $\omega_{\rm HDI}$ $\le$ $\omega_{\rm Goal}$ & Precision then Location \\
        Decisive PitG & $\omega_{\rm Goal}$ & $\omega_{\rm HDI}$ $\le$ $\omega_{\rm Goal}$ AND & Precision $+$ Location \\
         & & Decision is conclusive & \\
      \end{tabular}
      \caption{Comparison of stopping algorithms. `Decision is conclusive' means the HDI
      is fully inside or fully outside the ROPE. $\omega_{\rm HDI}$ denotes the observed
      width of the HDI and $\omega_{\rm Goal}$ denotes the target precision width.}
      \label{tab:comparison}
    \end{center}
  \end{table}

The Key Input column identifies the characteristic parameter that must be determined
before data collection: $N_{\rm min}$ for the HDI+ROPE algorithm (sensitive to early outliers and
therefore requiring a minimum sample size, e.g.\ 30) and $\omega_{\rm Goal}$ for the
precision methods, where by definition $\omega_{\rm Goal}$ must be narrower than or
equal to the ROPE width $\Delta_{\rm ROPE}$ to allow the HDI to fit fully within the ROPE.

The table also excludes two universal parameters required by all three algorithms:
\begin{enumerate}[label=(\alph*)]
    \item The ROPE boundaries $\mathrm{ROPE}_\mathrm{min}$ and $\mathrm{ROPE}_\mathrm{max}$,
    with $\Delta_\mathrm{ROPE} \equiv \mathrm{ROPE}_\mathrm{max} - \mathrm{ROPE}_\mathrm{min}$.
    \item $N_\mathrm{max}$ (Final Budget): The maximum sample size or budget limit.
\end{enumerate}

All three algorithms below assume one data point is collected per iteration
(Algorithms \ref{alg:hdi_rope}--\ref{alg:dpitg});
the increment step ($N \mathrel{+}= 1$) can be replaced by a batch size to
accommodate grouped collection.

\subsubsection{HDI+ROPE Stopping: Location Only}\label{sec:hdi_rope_stopping}

The HDI+ROPE algorithm stops data collection as soon as the posterior's location allows for a decisive acceptance or rejection of the null hypothesis. It does not explicitly demand a specific degree of precision (width), only that the HDI does not straddle the ROPE boundary.

Algorithm \ref{alg:hdi_rope} details the procedure. Note the inclusion of $N_\mathrm{min}$
(minimal sample size); this is critical because, with small sample sizes, the HDI can
fluctuate wildly and accidentally fall entirely within or outside the ROPE due to random noise
(aleatoric uncertainty)\footnote{Note the distinction from the epistemic uncertainty discussed
earlier: the HDI \textit{width} $\omega_{\rm HDI}$ is epistemic: it reflects how much we do not yet know about
$\theta$ and shrinks as $N$ grows. The HDI \textit{location}, however, varies across experiments
due to the randomness of which particular draws are observed; at small $N$ this aleatoric
variability dominates, and the HDI can wander into a misleadingly decisive position.} rather
than true effect.

\begin{algorithm}
    \caption{HDI+ROPE algorithm (highlighted: the stopping condition)}\label{alg:hdi_rope}
    \begin{algorithmic}
    \Require $N_\mathrm{min}$, $N_\mathrm{max}$, $\theta_{\rm null}$, $\mathrm{ROPE}_\mathrm{min}$, $\mathrm{ROPE}_\mathrm{max}$
    \State Stop = False
    \State Decision = Inconclusive
    \State N = 0
    \While{(Stop = False \& $N < N_\mathrm{max}$)} 
    \State N += 1 \Comment{collect another data point}
    \State Update $\hat\theta$, $P(\theta|\hat\theta)$ \Comment{update posterior}
    \State $\mathrm{HDI}_\mathrm{min}, \ \mathrm{HDI}_\mathrm{max}  \gets P(\theta|\hat\theta)$
    \If{$N \ge N_\mathrm{min}$}
        \State Decision = Decision Algorithm($\mathrm{ROPE}_\mathrm{min}, \mathrm{ROPE}_\mathrm{max}, \mathrm{HDI}_\mathrm{min}, \mathrm{HDI}_\mathrm{max}$)
        \If{Decision in \{Accept $\theta_{\rm null}$, Reject $\theta_{\rm null}$\}}
            \State \HiLi Stop = True \Comment{Stopping depends on decisive Decision}
        \EndIf
    \EndIf
    \EndWhile
    \end{algorithmic}
\end{algorithm}

This method serves as the coupled baseline throughout: lacking a precision requirement,
it is susceptible to false-reject and false-accept decisions, as demonstrated in
Section~\ref{sec:results}.

\subsubsection{Precision is the Goal Stopping: Precision then Decision}\label{sec:pitg}

Introduced by \cite{kruschke2015doing}, Precision is the Goal shifts the stopping condition
from the \textit{decision} to the \textit{precision} (width) of the estimate.
Data collection stops when the width of the HDI is narrower than a pre-specified Goal
(e.g., $\omega_{\rm Goal}=0.8 \Delta_{\rm ROPE}$).

Algorithm \ref{alg:pitg} shows that this method separates stopping from decision:
stopping depends on width alone, while the decision depends on location.
As we later demonstrate in Section \ref{sec:results}, this can lead to situations where
the algorithm stops (precision reached) but the result is inconclusive (posterior straddles the ROPE).

\begin{algorithm}
    \caption{Precision is the Goal (highlighted: the stopping condition)}\label{alg:pitg}
    \begin{algorithmic}
    \Require $\omega_\mathrm{Goal}$, $N_\mathrm{max}$, $\theta_{\rm null}$, $\mathrm{ROPE}_\mathrm{min}$, $\mathrm{ROPE}_\mathrm{max}$
    \State Stop = False
    \State Decision = Inconclusive
    \State N = 0
    \While{(Stop = False \& $N < N_\mathrm{max}$)}
    \State N += 1 \Comment{collect another data point}
    \State Update $\hat\theta, P(\theta|\hat\theta)$ \Comment{update posterior}
    \State $\mathrm{HDI}_\mathrm{min}, \mathrm{HDI}_\mathrm{max} \gets P(\theta|\hat\theta)$
    \State $\omega_{\rm HDI} = \mathrm{HDI}_\mathrm{max} - \mathrm{HDI}_\mathrm{min}$
    \If{$\omega_{\rm HDI} \le$ $\omega_\mathrm{Goal}$}
         \State \HiLi Stop = True \Comment{Stopping depends on Precision $\omega_{\rm HDI}$ regardless of Decision}
         \State Decision = Decision Algorithm($\mathrm{ROPE}_\mathrm{min}, \mathrm{ROPE}_\mathrm{max}, \mathrm{HDI}_\mathrm{min}, \mathrm{HDI}_\mathrm{max}$)  
    \EndIf
    \EndWhile
    \end{algorithmic}
\end{algorithm}

\subsubsection{Decisive Precision is the Goal Stopping: Precision and Decision}\label{sec:decisive_pitg}

We propose Decisive Precision is the Goal, a conservative variant of PitG,
to mitigate the high rate of inconclusive results. Unlike PitG, this method joins stopping
and decision: it stops strictly when \textit{both} the precision goal is met \textit{AND}
a decisive decision can be made.

As seen in Algorithm \ref{alg:dpitg}, strictly enforcing both conditions ensures that we do not stop in an indecisive state, potentially at the cost of collecting more data (up to $N_\mathrm{max}$).

\begin{algorithm}
    \caption{Decisive Precision is the Goal (highlighted: the stopping condition)}\label{alg:dpitg}
    \begin{algorithmic}
    \Require $\omega_\mathrm{Goal}$, $N_\mathrm{max}$, $\theta_{\rm null}$, $\mathrm{ROPE}_\mathrm{min}$, $\mathrm{ROPE}_\mathrm{max}$
    \State Stop = False
    \State Decision = Inconclusive
    \State N = 0
    \While{(Stop = False \& $N < N_\mathrm{max}$)}
    \State N += 1 \Comment{collect another data point}
    \State Update $\hat\theta, P(\theta|\hat\theta)$ \Comment{update posterior}
    \State $\mathrm{HDI}_\mathrm{min}, \ \mathrm{HDI}_\mathrm{max}  \gets P(\theta|\hat\theta)$ 
    \State $\omega_{\rm HDI} = \mathrm{HDI}_\mathrm{max} - \mathrm{HDI}_\mathrm{min}$
    \If{$\omega_{\rm HDI} \le$ $\omega_\mathrm{Goal}$}
        \State Decision = Decision Algorithm($\mathrm{ROPE}_\mathrm{min}, \mathrm{ROPE}_\mathrm{max}, \mathrm{HDI}_\mathrm{min}, \mathrm{HDI}_\mathrm{max}$) 
        \If{Decision in \{Accept $\theta_{\rm null}$, Reject $\theta_{\rm null}$\}} 
            \State \HiLi Stop = True \Comment{Stopping depends on Precision $\omega_{\rm HDI}$ and Decision}
        \EndIf
    \EndIf
    \EndWhile
    \end{algorithmic}
\end{algorithm}

\subsection{Expected Precision-Based Stop Iteration}\label{sec:expected_stop_iteration}

\begin{figure}[h!]
  \centering
  \includegraphics[width=1\textwidth]{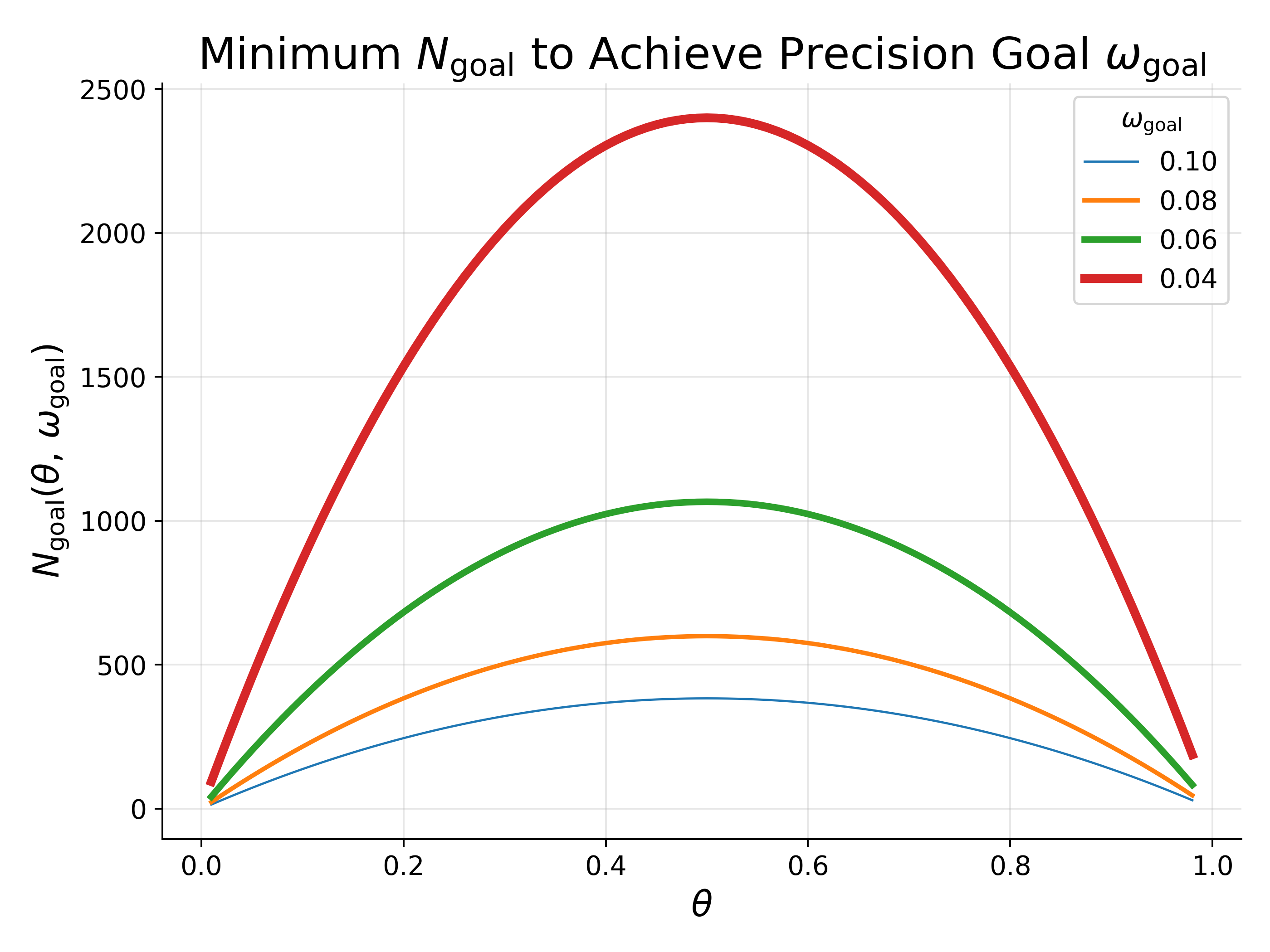}
  \caption{Expected minimum precision-based stop iteration
  $N_{\rm goal}(\theta, \omega_{\rm goal})$ as a function of
  $\theta$ for several values of $\omega_{\rm goal}$ in the Bernoulli case.
  (Equation~\ref{eq:pitg_stop_iteration})
  }
  \label{fig:min_sample_by_goal}
\end{figure}

For models where the posterior concentrates around the true parameter as data accumulate,
a property of all regular parametric families, guaranteed asymptotically by the
Bernstein--von Mises theorem \citep{vandervaart1998}, the HDI width shrinks
proportionally to $N^{-1/2}$.
This makes the expected minimum stopping iteration of precision-based algorithms predictable.
Under a normal approximation (central limit theorem; CLT), the expected sample size to achieve a precision goal
$\omega_{\rm goal}$ is proportional to the per-observation variance $V(\theta)$ of the
estimator (see Appendix~\ref{app:expected_stop} for the full derivation).
As a concrete example, for a Bernoulli rate parameter $\theta$ as in the setting introduced
in Section~\ref{sec:experimental_design}, $V(\theta) \propto \theta(1-\theta)$, giving
\begin{equation}\label{eq:pitg_stop_iteration}
N_{\rm goal}(\theta,\omega_{\rm goal}) = \frac{4 z_{*}^{2}}{\omega_{\rm goal}^{2}}\,\theta(1-\theta) - 1,
\end{equation}
where $z_{*}$ is the critical value for the chosen credibility level (e.g.,
$z_{*}\approx 1.96$ for 95\%). 
Because $\theta(1-\theta)$ is maximised at $\theta=0.5$, the fair-coin scenario demands
the largest sample for any given precision goal, and $N_{\rm goal}$ decreases monotonically
as $\theta$ approaches $0$ or $1$.
Figure~\ref{fig:min_sample_by_goal} illustrates this dependence
for the Bernoulli case across $\omega_{\rm goal}$.

This also illuminates DPitG's relationship to the HDI+ROPE algorithm: because the
precision condition must be satisfied before any decision is possible, DPitG behaves
approximately as the HDI+ROPE algorithm with $N_{\rm min} \approx N_{\rm goal}$, replacing
the ad-hoc floor (e.g.\ $N_{\rm min}=30$) with a precision-derived bound.

For completeness, we mention that for continuous outcomes, $V(\theta)$ is replaced by the
estimand variance $\sigma^2$; see Appendix~\ref{app:expected_stop} for the full
derivation, including a Student-$t$ variant for small samples.

\subsection{Experimental Design}\label{sec:experimental_design}

To evaluate and compare the performance of the three algorithms,
we employ a synthetic data simulation based on Bernoulli trials (dichotomous data).
While these methods are applicable to continuous data, we focus on the Bernoulli case
to maintain comparability with the work of \cite{kruschke2015doing}.\footnote{We follow
the setup of \S13.3 of \cite{kruschke2015doing} and expand upon it with a broader
range of $\theta_{\rm true}$ scenarios, systematic performance metrics across $M=5{,}000$
simulations, and visualisations designed to directly compare all three algorithms.}

We generate $M=5,000$ independent sequences of $N_{\rm max}=1,500$ trials each.
This choice exceeds $N_{\rm goal}$ at the most demanding setting tested ($\omega_{\rm goal}=0.06$,
$\theta_{\rm true}=0.5$), where Equation~\ref{eq:pitg_stop_iteration} gives $N_{\rm goal}\approx 1{,}067$,
ensuring that DPitG can reach its stopping criterion in the vast majority of experiments.

We test multiple scenarios for the ground truth probability $0.5 \le \theta_{\rm true} \le 0.8$
against a null hypothesis of $\theta_{\rm null}=0.5$, including the fair-coin case
$\theta_{\rm true}=\theta_{\rm null}=0.5$.
\subsubsection*{Simulation Configuration}

Unless stated otherwise, we use the following default parameters, chosen to reflect realistic experimental conditions (e.g., opinion polling):

\begin{enumerate}[label=(\alph*)]
    \item ROPE: Considering an effect size of $\pm 0.05$ around the $\theta_{\rm null}$, we define $\mathrm{ROPE}_{\rm max}^{\rm min}=[0.45,\,0.55]$. ($\Delta_{\rm ROPE}=0.1$)
    \item Precision Goal $\omega_{\rm goal}=0.08$: I.e, we set $\omega_{\rm goal} = 0.8 \cdot \Delta_{\rm ROPE}$.
    \item \textbf{Minimum Sample Size ($N_{\rm min}=30$)}: A practical lower bound to guard against misleadingly decisive stops at small $N$, where aleatoric variability can drive the HDI into or out of the ROPE before sufficient data have accumulated (see Section~\ref{sec:hdi_rope_stopping}).
    This parameter is required for the HDI+ROPE algorithm but not the precision-based algorithms.
    \item \textbf{Maximum Sample Size ($N_{\rm max}=1,500$)}: An upper bound for budgetary considerations. Experiments that are not decisive by this iteration are flagged as \textit{indecisive}.
\end{enumerate}

In the following section, we present the results of these simulations, first under these default settings and subsequently by varying key parameters to explore robustness.

\section{Results}\label{sec:results}
We present results in two parts.
We open with the fair coin case ($\theta_{\rm true}=\theta_{\rm null}=0.5$):
a hand-picked experiment to illustrate the qualitative behaviour of all three algorithms
(Section~\ref{sec:fair_coin_sequence}), followed by a large-scale study across
$M=5{,}000$ independent experiments (Section~\ref{sec:fair_coin_large_scale}) and a
sensitivity analysis over the precision goal $\omega_{\rm goal}$
(Section~\ref{sec:goal_impact}).
We then examine how all three algorithms perform as $\theta_{\rm true}$ increases from
$0.5$ upwards, across and beyond the ROPE (Section~\ref{sec:trends}).

\subsection{Fair Coin ($\theta_{\rm true}=0.5$)}\label{sec:fair_coin}

\subsubsection{A Hand-Picked Sequence}\label{sec:fair_coin_sequence}

\begin{figure}[h!]
  \centering
  \includegraphics[width=1\textwidth]{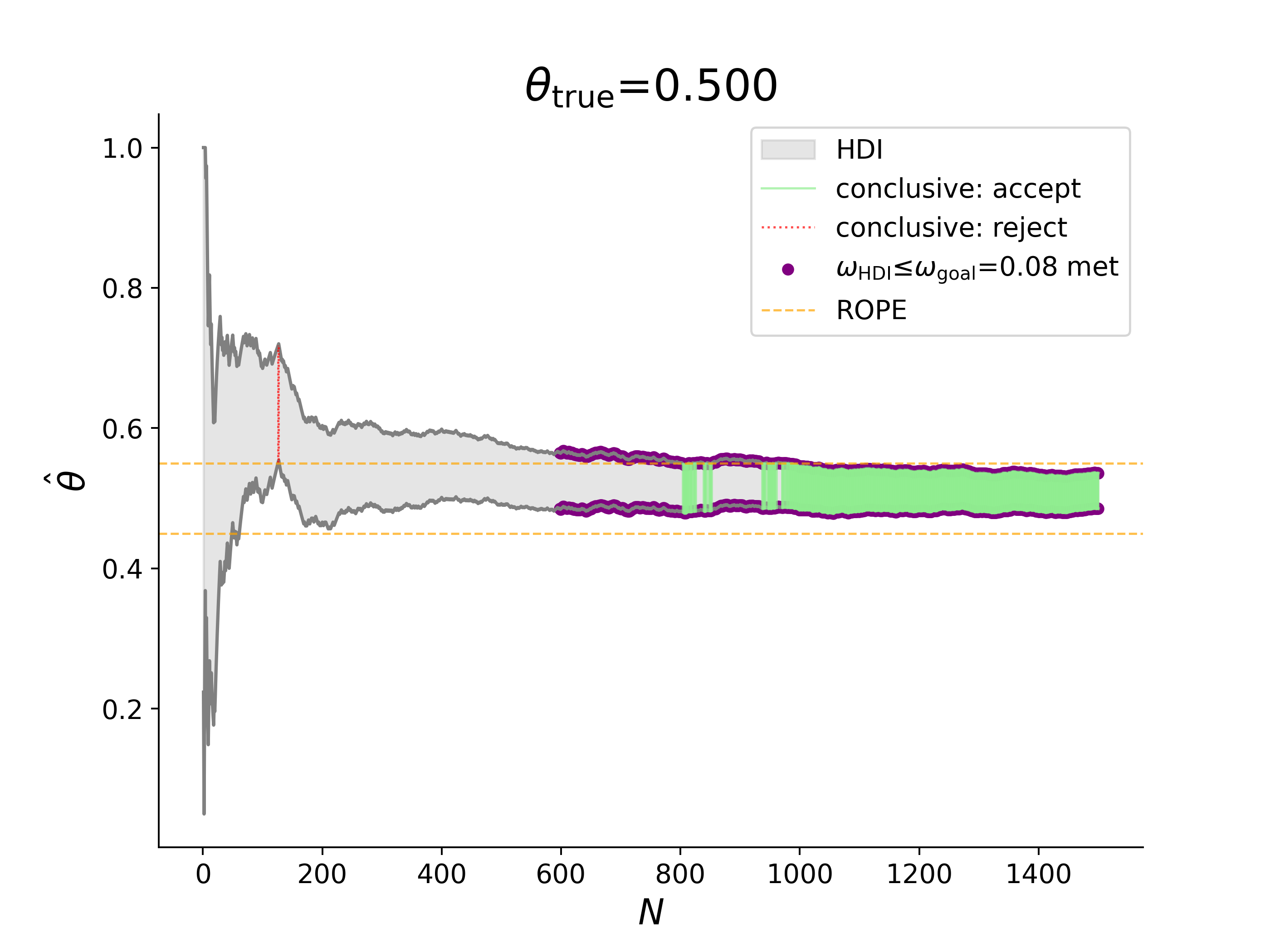}
  \caption{Cumulative sample proportion $\hat{\theta}(N)$ at each iteration $N$ of a
  hand-picked randomly generated fair coin experiment
  ($\theta_{\rm true}=\theta_{\rm null}=0.5$, $\omega_{\rm goal}=0.08$, $N_{\rm min}=30$)
  illustrating varied outcomes across sampling.
  $\hat{\theta}(N)$ denotes the observed proportion of successes (heads) in $N$ trials.
  \textit{Gray band}: 95\% HDI of the posterior; its width is $\omega_{\rm HDI}$.
  \textit{Horizontal dashed lines}: ROPE boundaries.
  \textit{Vertical red dashed line}: Decision to reject $\theta_{\rm null}$.
  \textit{Vertical green solid line}: Decision to accept $\theta_{\rm null}$.
  \textit{Purple dots}: Precision goal met ($\omega_{\rm HDI} \le \omega_{\rm goal}$).
  See Figure~\ref{fig:posteriors} for the posterior distributions at the three marked
  stop iterations.}
  \label{fig:iterations}
\end{figure}

Figure~\ref{fig:iterations} presents a hand-picked randomly generated fair coin experiment chosen to
illustrate contrasting outcomes across the three algorithms.\footnote{This figure is similar to the top
panels in \cite{kruschke2015doing} Figures 13.4 and 13.5.}
(note that all performance claims rest on the large-scale simulation study of
Section~\ref{sec:fair_coin_large_scale})
The posteriors in Figure~\ref{fig:posteriors} correspond to three specific iterations
of this same experiment.\footnote{The sequence may be found in the open-source material
as \href{https://github.com/elzurdo/precision-goal/blob/main/py/utils_experiments.py\#L28}{\texttt{SEQUENCE\_HANDPICKED}}.}

Despite arising from the same sequence of coin tosses, the three algorithms reach
sharply different decisions.
The HDI+ROPE algorithm stops early at iteration 126 (vertical red dashed line) and, incorrectly,
rejects $\theta_{\rm null}$.
PitG stops at iteration 598 (first purple dot, where the precision goal is first met)
but is left inconclusive, as the HDI straddles the ROPE boundary.
DPitG waits until both conditions are jointly satisfied, stopping at iteration 804
(first green solid line coinciding with a purple dot) and correctly accepts
$\theta_{\rm null}$.
This divergence from a single experiment motivates the large-scale analysis that
follows.

\subsubsection{Large-Scale Experiments}\label{sec:fair_coin_large_scale}

\begin{figure}[h!]
  \centering
  \includegraphics[width=1\textwidth]{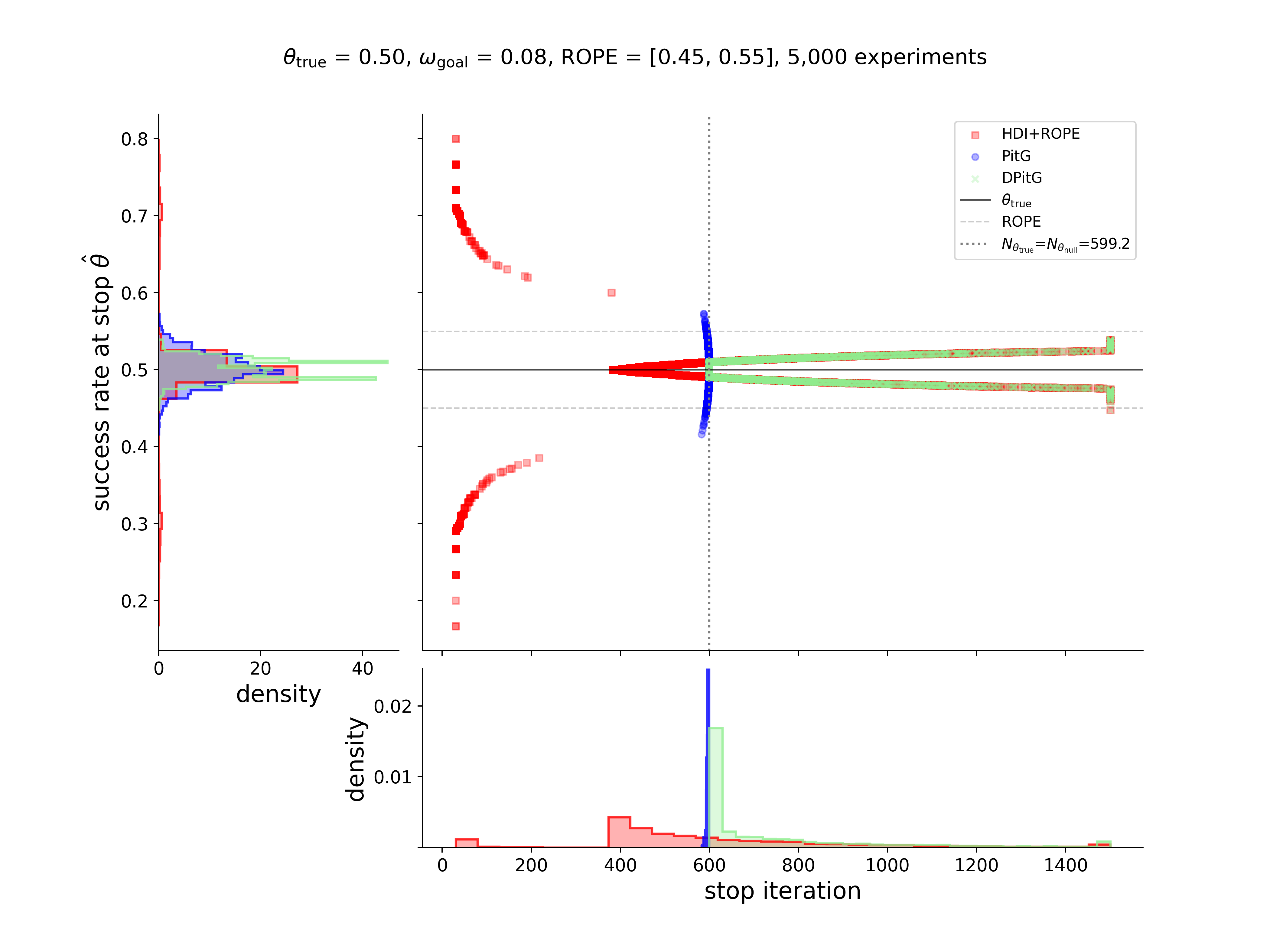}
  \caption{Outcomes of $M=5{,}000$ fair coin experiments ($\theta_{\rm true}=\theta_{\rm null}=0.5$,
  $\omega_{\rm goal}=0.08$, $N_{\rm min}=30$).
  \textit{Main panel}: scatter plot of sample proportion $\hat{\theta}(N_{\rm stop})$
  vs.\ stop iteration $N_{\rm stop}$.
  \textit{Symbols}: red squares: HDI+ROPE; blue circles: PitG; green Xs: DPitG.
  \textit{Solid horizontal line}: $\theta_{\rm true}$.
  \textit{Horizontal dashed lines}: ROPE boundaries.
  \textit{Vertical dashed line}: expected goal stop iteration $N_{\rm goal}$
  (Equation~\ref{eq:pitg_stop_iteration}); here $N_{\rm goal}(\theta_{\rm true})=
  N_{\rm goal}(\theta_{\rm null})$ since $\theta_{\rm true}=\theta_{\rm null}$.
  \textit{Bottom panel}: histograms of $N_{\rm stop}$.
  \textit{Left panel}: histograms of $\hat{\theta}(N_{\rm stop})$.
  The marginal histograms reveal where most results concentrate,
  complementing the scatter plot which shows the correlations.
  Histogram binning varies by algorithm to best resolve each distribution.}
  \label{fig:fair_iter_vs_rate}
\end{figure}

\begin{figure}[h!]
  \centering
  \includegraphics[width=1\textwidth]{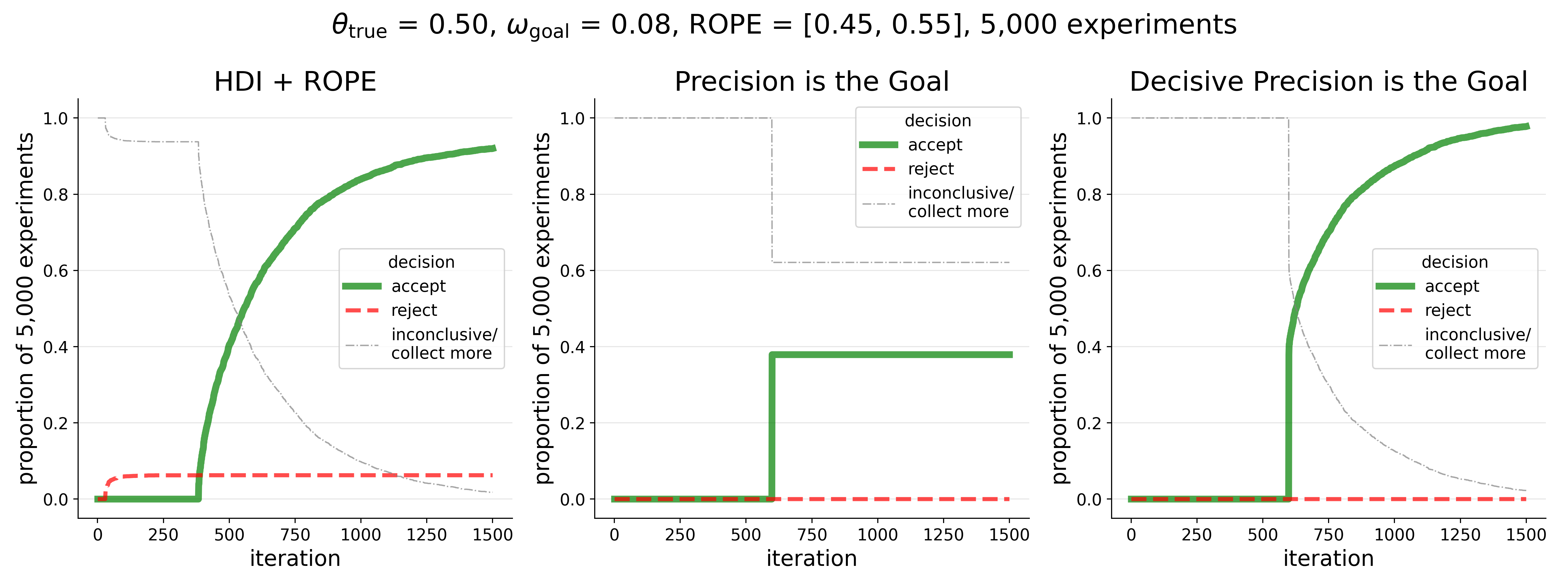}
  \caption{Cumulative decision rates across iterations for $M=5{,}000$ fair coin experiments
  ($\theta_{\rm true}=\theta_{\rm null}=0.5$, $\omega_{\rm goal}=0.08$, $N_{\rm min}=30$).
  At each iteration $N$, the plotted proportions count all experiments that reached
  a decision at or before $N$; their sum is 100\% at every point.
  \textit{Left}: HDI+ROPE; \textit{Middle}: Precision is the Goal; \textit{Right}: Decisive Precision is the Goal.
  \textit{Red dashed line}: reject $\theta_{\rm null}$.
  \textit{Green solid line}: accept $\theta_{\rm null}$.
  \textit{Gray dot-dashed line}: inconclusive.}
  \label{fig:fair_decisions}
\end{figure}

We expand this analysis to $M=5{,}000$ experiments, displaying outcomes in
Figures~\ref{fig:fair_iter_vs_rate} and~\ref{fig:fair_decisions},\footnote{Both of these
visuals are inspired by Figure~13.6 of \cite{kruschke2015doing}.} and summarising key
statistics in Tables~\ref{tab:fair_overall} and~\ref{tab:fair_conclusive}.

Figure~\ref{fig:fair_iter_vs_rate} shows that the HDI+ROPE algorithm yields a wide range of stop
iterations and sample proportions $\hat{\theta}$, with a non-negligible proportion
of results falling outside the ROPE ($\hat{\theta}$ ranging from $\sim\!0.25$ to
$\sim\!0.80$), leading to incorrect rejections
of $\theta_{\rm null}$. Its $N_{\rm stop}$ distribution is skewed towards early stopping
at $\sim\!400$, well before $N_{\rm goal}\sim 600$, with a long tail extending to later
iterations.

PitG consistently stops near $N_{\rm goal}$, with sample proportions clustered tightly around
$\theta_{\rm true}=0.5$. However, many of these stops occur when the HDI straddles the
ROPE boundary, yielding a high rate of inconclusive decisions
(see Figure~\ref{fig:fair_decisions} and Table~\ref{tab:fair_overall}).
DPitG exhibits a narrower $\hat{\theta}(N_{\rm stop})$ distribution than PitG but extends
to later iterations, achieving more conclusive results while maintaining zero estimation bias.\footnote{
In Section~\ref{sec:trends} we demonstrate that some bias emerges as symmetry breaks.}
The $N_{\rm stop}$ histograms in Figure~\ref{fig:fair_iter_vs_rate} make this concrete:
DPitG's distribution begins near $N_{\rm goal}$, precisely where PitG's terminates,
confirming the structural equivalence noted in
Section~\ref{sec:expected_stop_iteration}: DPitG behaves as the HDI+ROPE algorithm with
$N_{\rm min} \approx N_{\rm goal}$ rather than a fixed small floor.

Figure~\ref{fig:fair_decisions} shows that the HDI+ROPE algorithm's rejection rate (left panel) quickly asymptotes to 6\%, while its acceptance
rate begins rising near $N\sim 400$, reflecting a gradual shift from incorrect rejections
to correct acceptances as more data accumulate.

Both precision-based methods maintain a zero rejection rate (middle and right panels),
as they require the precision criterion to be satisfied before any decision is made.
For PitG (middle panel), the acceptance rate rises as a near step-function, jumping at
$N_{\rm goal}$ to 37.9\%, with the inconclusive rate settling at the complement 62.1\%.

DPitG (right panel) presents a hybrid of the other two: it preserves the zero-rejection
property of PitG while achieving a high conclusiveness rate akin to the HDI+ROPE algorithm, but with
a more gradual transition from inconclusive to conclusive outcomes.

Table~\ref{tab:fair_overall} summarises stop-iteration statistics across all $M$ experiments;
Table~\ref{tab:fair_conclusive} then restricts to the conclusive subsets and breaks outcomes
into acceptance and rejection rates alongside the $\hat{\theta}$ distribution at stop.

\begin{table}[h!]
  \begin{center}
  \begin{tabular}{c|c|c|c|c}
    \hline
    Algorithm &  Conclusive Rate  & $N_{\rm 50\%}$ & $N_{\rm 25\%}$--$N_{\rm 75\%}$ & $\Delta_{\rm IQR}$\\
    \hline
    HDI+ROPE & 0.982 & 521 & 415 - 723& 308  \\
    PitG     & \textcolor{red}{\textbf{0.379}}  & 599 & 598 - 599& 1 \\
    DPitG    & \textcolor{green}{\textbf{0.978}}  & 627 & 598 - 794& 196 \\
    \hline
  \end{tabular}
  \caption{Summary statistics for all $M=5{,}000$ fair coin experiments
  ($\theta_{\rm true}=\theta_{\rm null}=0.5$, $\omega_{\rm goal}=0.08$, $N_{\rm min}=30$;
  cf.\ Figure~\ref{fig:fair_iter_vs_rate}).
  $N_{\rm 50\%}$: median stop iteration;
  $N_{\rm 25\%}$--$N_{\rm 75\%}$: interquartile range of stop iteration.
  Conclusive Rate values are color-coded: \textcolor{red}{\textbf{red}} flags a
  concerningly low rate (PitG); \textcolor{green}{\textbf{green}} highlights DPitG's
  strong rate with zero false rejections (cf.\ Table~\ref{tab:fair_conclusive}).}
  \label{tab:fair_overall}
\end{center}
\end{table}

The HDI+ROPE algorithm achieves the highest conclusiveness rate (98.2\%), but at the cost of a 6.3\%
false-reject rate. PitG delivers zero false rejections, but its conclusiveness rate of
only 37.9\% is strikingly low, with stop iterations all clustering at $N_{\rm goal}$.
DPitG delivers the balance sought in Section~\ref{sec:decisive_pitg}: a 97.8\%
conclusiveness rate with zero false
rejections. The lower quartile ($N_{\rm 25\%}=598$) nearly equals $N_{\rm goal}=599$,
confirming that a substantial fraction of DPitG experiments stop as soon as the
precision goal is first met; the wider IQR of $598$--$794$ reflects those that
require additional iterations to satisfy the joint precision-and-conclusiveness
criterion, at a median cost of only 4.7\% more samples across all $M$ experiments
(4\% when restricted to conclusive experiments; see Table~\ref{tab:fair_conclusive})
than PitG.

\begin{table}[h!]
  \begin{center}
  \begin{tabular}{c|c|c|c|c|c}
    \hline
    Algorithm & Conclusive & Acceptance Rate & Rejection Rate &  $\hat{\theta}_{\rm 50\%}$ & $\hat{\theta}_{\rm 25\%} - \hat{\theta}_{\rm 75\%}$\\
    \hline
    HDI+ROPE & 4912 & 0.936 & \textcolor{red}{\textbf{0.064}} & 0.5000 & 0.4920-0.5081 \\
    PitG     &  \textbf{1895} & 1     & 0     & 0.5008 & 0.4958-0.5041 \\
    DPitG    & 4888 & 1     & 0     & 0.5008 & 0.4895-0.5105\\
    \hline
  \end{tabular}
  \caption{Summary statistics for the conclusive subsets of experiments in
  Table~\ref{tab:fair_overall}.
  Conclusive: count of experiments (out of 5{,}000) with a conclusive outcome.
  Acceptance and Rejection rates are conditioned on conclusive experiments
  and therefore sum to 1.
  $\hat{\theta}_{\rm 50\%}$: median sample proportion at stop;
  $\hat{\theta}_{\rm 25\%} - \hat{\theta}_{\rm 75\%}$: interquartile range of
  sample proportion at stop.
  \textcolor{red}{\textbf{Red}} flags a non-zero false-reject rate (HDI+ROPE); \textbf{bold} flags a concerningly low conclusive count (PitG).}
  \label{tab:fair_conclusive}
\end{center}
\end{table}

\subsubsection{Sensitivity to $\omega_{\rm goal}$}\label{sec:goal_impact}

The choice of $\omega_{\rm goal}$ governs the fundamental precision--conclusiveness
trade-off for PitG. Table~\ref{tab:goal_impact} shows the effect across a range of values
at $\theta_{\rm true}=0.5$. Tightening $\omega_{\rm goal}$ (decreasing it) raises PitG's
conclusiveness at the cost of a larger required sample size: $N_{\rm goal}$ grows
quadratically as $\omega_{\rm goal}$ decreases (Equation~\ref{eq:pitg_stop_iteration}).
DPitG's conclusiveness, by contrast, remains essentially flat across the full range
($0.973$--$0.980$), demonstrating robustness to the choice of $\omega_{\rm goal}$.

The conclusiveness ratio (DPitG/PitG) captures this contrast directly: it grows from
$1.19$ at $\omega_{\rm goal}=0.06$ to undefined at $\omega_{\rm goal}=0.10$, where PitG
becomes entirely inconclusive. DPitG's robustness comes at the cost of larger sample
sizes, quantified by the $N_{\rm 50\%}/N_{\rm goal}$ column: for $\omega_{\rm goal} \le 0.07$,
the median DPitG stop equals $N_{\rm goal}$ (no premium); for looser goals the median
sampling premium rises from 4\% (cf.\ 4.7\% across all $M$ experiments) at
$\omega_{\rm goal}=0.08$ to 41\% at $\omega_{\rm goal}=0.10$.

At the tightest setting ($\omega_{\rm goal}=0.06$), the DPitG $N_{\rm stop}$ $\Delta_{\rm IQR}$ is zero:
most experiments stop at exactly $N_{\rm goal}$, leaving no room for the additional
iterations that DPitG normally uses to resolve borderline cases.
At the loosest setting, when equal to the ROPE width ($\omega_{\rm goal}=0.10=\Delta_{\rm ROPE}$), PitG is always
inconclusive, but DPitG still achieves 97.9\% conclusiveness by collecting more data until
both precision and conclusiveness are jointly satisfied.

The conclusiveness of DPitG is more sensitive to $N_{\rm max}$ (fixed at $1{,}500$ in this
example) than to $\omega_{\rm goal}$, whereas PitG always stops at $N_{\rm goal}$ regardless of $N_{\rm max}$.
Consequently, $N_{\rm goal}$ serves as a minimum budget for DPitG: for tight precision goals
it is sufficient to achieve high conclusiveness; for looser goals, the researcher should
set $N_{\rm max}$ generously to allow DPitG to reach its full potential.

\begin{table}[h!]
  \begin{center}
  \begin{tabular}{c|c|c|c|c|c|c|c|c|c}
    \hline
    & & \multicolumn{3}{c|}{Conclusiveness} & \multicolumn{5}{c}{DPitG $N_{\rm stop}$, conclusive experiments} \\
    $\omega_{\rm goal}$ & $N_{\rm goal}$ & PitG & DPitG & Ratio (DPitG/PitG) & $N_{\rm 25\%}$ & $N_{\rm 75\%}$ & $\Delta_{\rm IQR}$ & $N_{\rm 50\%}$ & $N_{\rm 50\%}/N_{\rm goal}$\\
    \hline
    0.06 & 1{,}066 & 0.818 & 0.973 & 1.19 & 1{,}066 & 1{,}066 &   0 & 1{,}066 & 1.00\\
    0.07 &   783   & 0.599 & 0.976 & 1.63 &   783   &   830   &  47 &   783   & 1.00\\
    0.08 &   599   & 0.379 & 0.978 & 2.58 &   599   &   779   & 180 &   623   & 1.04\\
    0.09 &   473   & 0.137 & 0.979 & 7.15 &   481   &   755   & 274 &   568   & 1.20\\
    0.10 &   383   & 0.000 & 0.980 & ---  &   428   &   734   & 306 &   538   & 1.41\\
    \hline
  \end{tabular}
  \caption{Impact of $\omega_{\rm goal}$ on PitG and DPitG performance
  ($\theta_{\rm true}=\theta_{\rm null}=0.5$, $M=5{,}000$ experiments,
  $N_{\rm max}=1{,}500$, $\Delta_{\rm ROPE}=0.1$).
  $N_{\rm goal}$: expected precision-based stop iteration (Equation~\ref{eq:pitg_stop_iteration});
  Conclusiveness: fraction of $M$ experiments with a conclusive outcome;
  Conclusiveness Ratio: DPitG/PitG conclusiveness (undefined at $\omega_{\rm goal}=0.10$
  where PitG conclusiveness is zero);
  DPitG $N_{\rm stop}$: quartile statistics restricted to conclusive DPitG experiments
  (results for $\omega_{\rm goal}=0.08$ differ slightly from Table~\ref{tab:fair_overall},
  which summarises all experiments);
  $\Delta_{\rm IQR} = N_{\rm 75\%} - N_{\rm 25\%}$.
  For $\omega_{\rm goal} \le 0.08$, $N_{\rm 25\%} = N_{\rm goal}$, showing that at least
  a quarter of DPitG experiments stop as soon as the precision goal is first met; this is
  also reflected in the $N_{\rm 50\%}/N_{\rm goal}$ column, which equals 1.00 for
  $\omega_{\rm goal} \le 0.07$.}
  \label{tab:goal_impact}
\end{center}
\end{table}

The fair coin case thus demonstrates that DPitG combines the HDI+ROPE algorithm's high
conclusiveness with PitG's zero-false-reject guarantee at a modest median cost.
Section~\ref{sec:trends} shows that this advantage persists as $\theta_{\rm true}$
departs from $\theta_{\rm null}$.

\subsection{General Trends Across $\theta_{\rm true}$ Values}\label{sec:trends}

\begin{figure}[h!]
  \centering
  \includegraphics[width=1\textwidth]{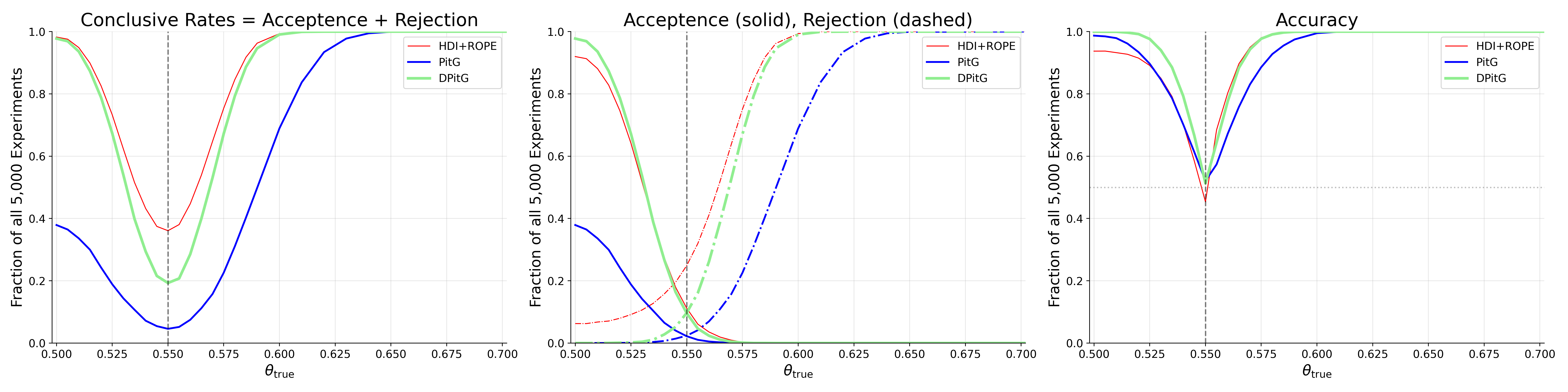}
  \caption{Algorithm comparison across $\theta_{\rm true}$ values for $M=5{,}000$
  experiments ($\theta_{\rm null}=0.5$, $\omega_{\rm goal}=0.08$).
  \textit{Colors}: red: HDI+ROPE; blue: PitG; green: DPitG.
  \textit{Vertical dashed lines}: ${\rm ROPE}_{\rm max}=0.55$.
  \textit{Left}: conclusiveness rates.
  \textit{Middle}: acceptance (solid lines) and rejection (dashed lines) rates;
  at each $\theta_{\rm true}$ their sum equals the respective conclusiveness rate.
  \textit{Right}: accuracy rates (proportion of correct decisions out of all
  $M$ experiments); inconclusive experiments are assigned a decision based on whether
  $\hat{\theta}$ falls inside or outside the ROPE.}
  \label{fig:conclusiveness_rates}
\end{figure}

\begin{figure}[h!]
  \centering
  \includegraphics[width=1\textwidth]{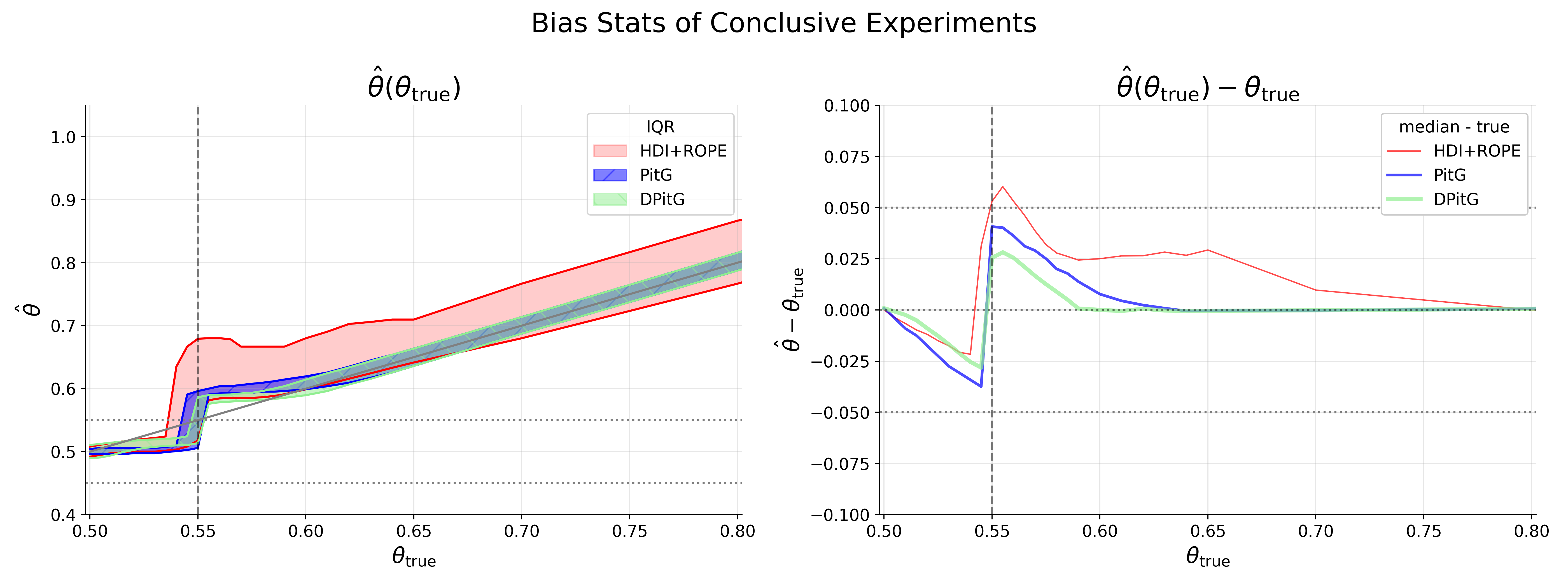}
  \caption{Sample proportion $\hat{\theta}(N_{\rm stop})$ at the stop iteration as a
  function of $\theta_{\rm true}$, for conclusive experiments only (subsets of $M=5{,}000$,
  $\theta_{\rm null}=0.5$, $\omega_{\rm goal}=0.08$, $N_{\rm min}=30$).
  \textit{Colors}: red: HDI+ROPE; blue: PitG; green: DPitG.
  \textit{Horizontal dotted lines}: ROPE boundaries.
  \textit{Vertical dashed line}: ${\rm ROPE}_{\rm max}$.
  \textit{Left}: interquartile range $\hat{\theta}_{\rm 25\%}$--$\hat{\theta}_{\rm 75\%}$
  of $\hat{\theta}(N_{\rm stop})$.
  \textit{Diagonal line}: parity line $\hat{\theta}=\theta_{\rm true}$.
  \textit{Right}: median bias $\hat{\theta}_{\rm 50\%}-\theta_{\rm true}$.}
  \label{fig:success_by_truth_conclusive}
\end{figure}

\begin{figure}[h!]
  \centering
  \includegraphics[width=1\textwidth]{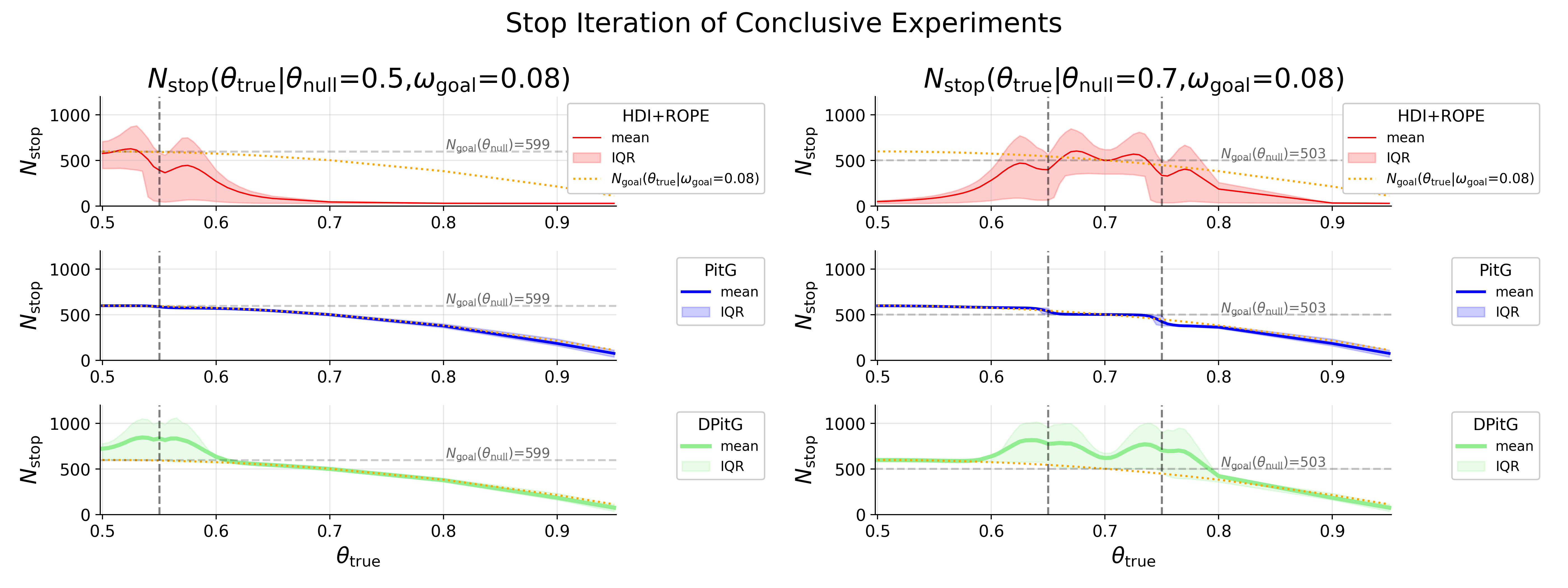}
  \caption{Stop iteration $N_{\rm stop}$ statistics for conclusive experiments across
  $\theta_{\rm true}$ values ($\omega_{\rm goal}=0.08$).
  \textit{Left}: $\theta_{\rm null}=0.5$; \textit{Right}: $\theta_{\rm null}=0.7$.
  In each column the top, middle, and bottom panels correspond to HDI+ROPE, PitG,
  and DPitG, respectively.
  \textit{Band}: interquartile range $N_{\rm 25\%}$--$N_{\rm 75\%}$ of $N_{\rm stop}$.
  \textit{Solid line}: median $N_{\rm 50\%}$.
  \textit{Dotted line}: true precision-goal iteration
  $N_{\rm goal}(\theta_{\rm true}, \omega_{\rm goal})$ (Equation~\ref{eq:pitg_stop_iteration}).
  \textit{Horizontal dashed line}: null precision-goal iteration
  $N_{\rm goal}(\theta_{\rm null}, \omega_{\rm goal})$, the researcher's planned sample size.}
  \label{fig:stop_iterations_by_truth_conclusive}
\end{figure}

\begin{figure}[h!]
  \centering
  \includegraphics[width=1\textwidth]{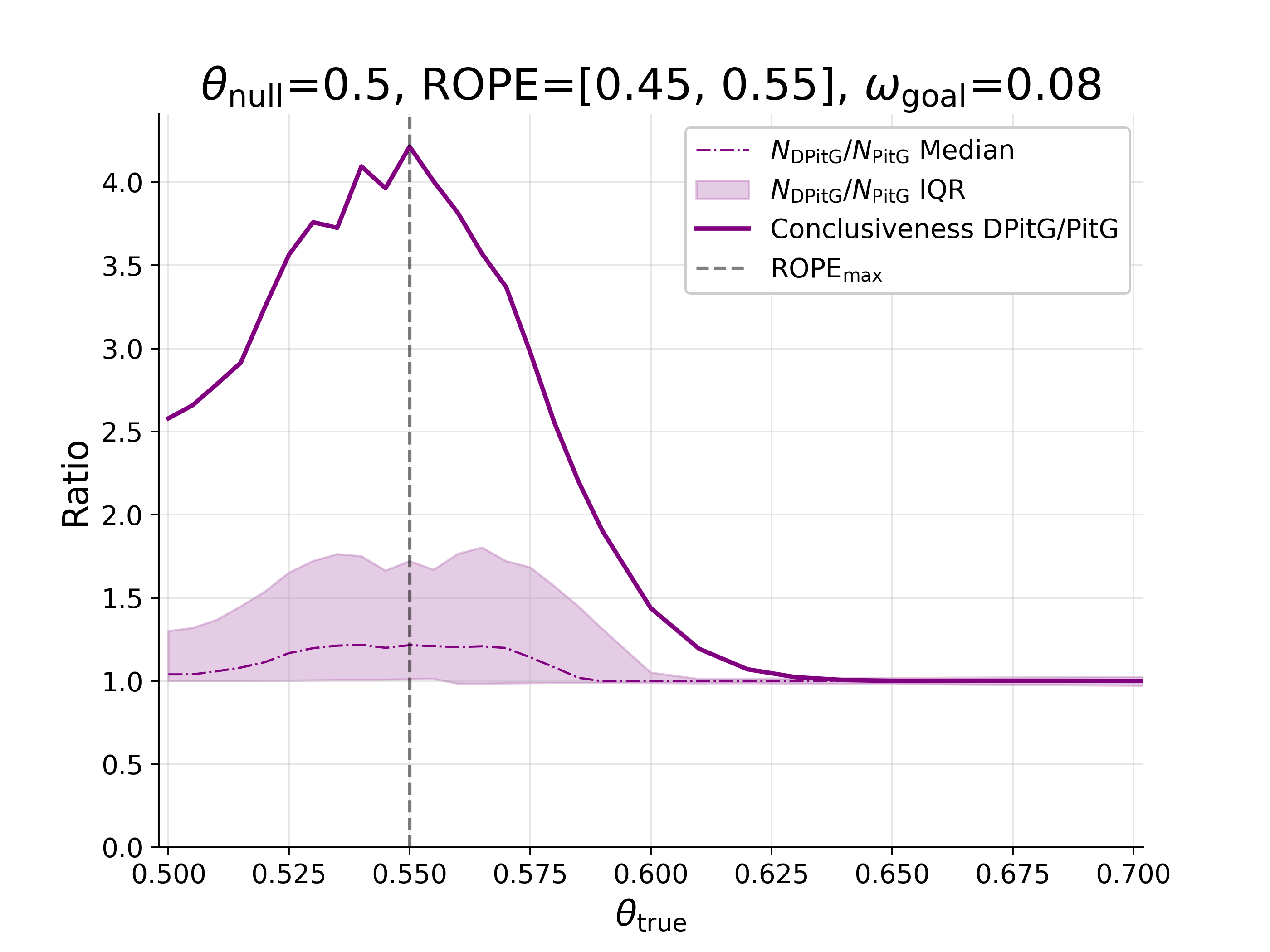}
  \caption{DPitG-to-PitG ratios of conclusiveness rate and stop iteration across
  $\theta_{\rm true}$ values ($M=5{,}000$, $\theta_{\rm null}=0.5$,
  $\omega_{\rm goal}=0.08$).
  \textit{Solid line}: conclusiveness ratio
  (cf.\ Figure~\ref{fig:conclusiveness_rates}).
  \textit{Dot-dashed line}: median $N_{\rm stop}$ ratio.
  \textit{Band}: interquartile range of the $N_{\rm stop}$ ratio.
  \textit{Vertical dashed line}: ${\rm ROPE}_{\rm max}=0.55$.}
  \label{fig:stop_conclusiveness_ratios}
\end{figure}

Building on the fair coin baseline, we now examine how all three algorithms perform as
$\theta_{\rm true}$ increases from $0.5$ upwards, across and beyond the ROPE,
\footnote{By symmetry, results for $\theta_{\rm true}<0.5$ mirror those shown
here, as reflected in Equation~\ref{eq:pitg_stop_iteration} and
Figure~\ref{fig:min_sample_by_goal}.}
with ${\rm ROPE}_{\rm max}=0.55$ serving as a key inflection point.
The primary case throughout is $\theta_{\rm null}=0.5$, where the null coincides
with the precision maximum; we also include $\theta_{\rm null}=0.7$ in
Figure~\ref{fig:stop_iterations_by_truth_conclusive} to examine behaviour when
this symmetry breaks and the planned sample size is no longer a conservative
upper bound.
We present four complementary views via
Figures~\ref{fig:conclusiveness_rates}--\ref{fig:stop_conclusiveness_ratios}:
conclusiveness and decision rates, overall accuracy, estimation bias at stop, and a
direct cost--benefit comparison of DPitG against PitG; qualitative trends generalise
across $\omega_{\rm goal}$ values (Section~\ref{sec:goal_impact}).

The left panel of Figure \ref{fig:conclusiveness_rates} shows that all three algorithms show a dip in conclusiveness as $\theta_{\rm true}$ approaches
${\rm ROPE}_{\rm max}$, after which conclusiveness rises again. The HDI+ROPE algorithm and DPitG
recover to near 100\% by $\theta_{\rm true}=0.6$; PitG only asymptotes to 100\% at
$\theta_{\rm true}\approx 0.64$, at which point DPitG and PitG become effectively
equivalent (for $\theta_{\rm null}=0.5$).

The precision-based methods are deliberately conservative near the ROPE boundaries,
yielding lower conclusiveness rates than the HDI+ROPE algorithm in this region; at ${\rm ROPE}_{\rm max}$
itself, DPitG is conclusive in only 19\% of experiments and PitG in only 5\%.
This dip is a natural consequence of the Decision Rule
(see Section~\ref{sec:accept_reject_meaning}): the ROPE boundaries are exactly the region
where neither an accept nor a reject verdict is easily justified.

Breaking the conclusiveness into its acceptance and rejection components (middle panel of Figure \ref{fig:conclusiveness_rates})
reveals the error structure.
For $\theta_{\rm null}=0.5$, acceptance is the correct decision within the ROPE
($\theta_{\rm true}\le 0.55$) and rejection is correct outside it.
The HDI+ROPE algorithm's false-reject rate climbs from $\sim\!6\%$ at $\theta_{\rm true}=0.5$
to $\sim\!30\%$ at ${\rm ROPE}_{\rm max}$; its false-accept rate peaks at
$\sim\!10\%$ at the ROPE boundary before dropping to $0\%$ by
$\theta_{\rm true}=0.57$. Both precision-based methods maintain a near-zero false-reject rate throughout:
exactly zero well within the ROPE interior, rising to a small level near the boundary:
DPitG reaching $\sim\!10\%$, and PitG lower only because its
conclusiveness rate is much smaller there.
Both also have false-accept rates that mirror the false-reject rates around
the ROPE boundary, which drops quickly as $\theta_{\rm true}$ moves outside.

The right panel of Figure \ref{fig:conclusiveness_rates} shows overall \textit{accuracy}, the proportion of correct decisions
across all $M$ experiments. Since the three algorithms have different conclusiveness
rates, inconclusive experiments must be assigned a decision for comparison.
We adopt the natural default: accept if $\hat{\theta}$ lies within the ROPE, reject
otherwise, the rule a practitioner would apply informally using the point estimate
alone, and the most common approach in empirical settings where credible intervals are
not examined; a more critical treatment of alternative imputation rules is deferred
to future work.

All three algorithms reach their minimum accuracy at ${\rm ROPE}_{\rm max}$. Within
the ROPE, DPitG consistently leads PitG, reaching 99.5\% at $\theta_{\rm true}=0.5$
compared to PitG's 98.5\%. Outside the ROPE, the HDI+ROPE algorithm and DPitG reach 100\% by
$\theta_{\rm true}=0.58$, whereas PitG achieves this only at
$\theta_{\rm true}\approx 0.60$.

That the HDI+ROPE algorithm reaches 100\% accuracy at $\theta_{\rm true}=0.58$ despite
producing biased $\hat{\theta}$ estimates motivates a closer look at the observed
sample proportion statistics. Figure~\ref{fig:success_by_truth_conclusive} shows, for
conclusive experiments, the IQR of $\hat{\theta}(N_{\rm stop})$ (left panel) and the
median bias $\hat{\theta}_{\rm 50\%}-\theta_{\rm true}$ (right panel).\footnote{We find
similar trends for the mean, but omit it to reduce clutter.}

As discussed in Section~\ref{sec:accept_reject_meaning}, accept/reject decisions rest on
the HDI's position relative to the ROPE, not on $\hat{\theta}$; a bias in $\hat{\theta}$
does not invalidate the decision so long as the HDI falls cleanly within (or outside)
the ROPE.

The HDI+ROPE algorithm's $\hat{\theta}$ bias is therefore more telling as a symptom of its lack of a
precision requirement than as a decision-quality metric. At $\theta_{\rm true}=0.5$,
all three algorithms show zero bias, as expected from the symmetry of the fair coin
case. Once $\theta_{\rm true}$ deviates from $0.5$, the HDI+ROPE algorithm's bias grows markedly: its
median bias approaches ${\rm ROPE}_{\rm max}$, with the upper
IQR extending substantially further, because the HDI+ROPE algorithm can stop on skewed samples before
sufficient data have accumulated to constrain the posterior.

The precision-based methods exhibit substantially smaller $\hat{\theta}$ bias, well
within the ROPE width across all $\theta_{\rm true}$ values, as a natural consequence
of their precision requirement. Among conclusive experiments within the ROPE, PitG
shows higher bias than DPitG; for inconclusive experiments (not shown\footnote{Bias
statistics for inconclusive experiments and the pooled analysis are available in the
open-source materials.}) (where PitG contributes far more cases than DPitG)
the pattern reverses, PitG's bias being slightly lower.
Pooled across both subsets, the aggregate difference is negligible: both methods yield
near-zero bias at all $\theta_{\rm true}$ values, with DPitG's aggregate bias only
marginally higher than PitG's.
Despite this, DPitG decision accuracy exceeds PitG's at every $\theta_{\rm true}$ value
(cf.\ Figure~\ref{fig:conclusiveness_rates}, right panel), because its substantially
higher conclusiveness rate more than compensates.

DPitG is decisively superior to PitG near and within the ROPE, at the cost of a larger
median $N_{\rm stop}$; having characterised the estimation properties of that superiority,
we now quantify its sampling cost.
Figure~\ref{fig:stop_iterations_by_truth_conclusive} quantifies $N_{\rm stop}$ for all three
algorithms, restricting to conclusive outcomes since inconclusive experiments always stop
at $N_{\rm max}$ and their stop iteration therefore carries no information about
algorithm behaviour; including the HDI+ROPE algorithm also illustrates why the precision
requirement matters.

The HDI+ROPE algorithm shows distinctly different behaviour inside and outside the ROPE.
Within the ROPE, $N_{\rm stop}$ loosely tracks $N_{\rm goal}(\theta_{\rm true})$, but
with a wide IQR reflecting the absence of a precision criterion.
Right at ${\rm ROPE}_{\rm max}$, a distinctive V-shaped signature is visible:
$N_{\rm stop}$ dips to a narrow trough at the boundary, as the few conclusive
experiments there stop early by catching a chance excursion of the HDI just outside
the ROPE. 
Outside the ROPE, $N_{\rm stop}$ falls well below $N_{\rm goal}(\theta_{\rm true})$:
The HDI+ROPE algorithm correctly
rejects $\theta_{\rm null}$ much faster than precision-based methods, but does so on
the basis of imprecise posteriors, producing the strong $\hat{\theta}$ bias
(cf.\ Figure~\ref{fig:success_by_truth_conclusive}).

PitG stops tightly around $N_{\rm goal}(\theta_{\rm true})$ by design, making it the
natural reference for understanding the precision cost. The contrast with the planned
quantity $N_{\rm goal}(\theta_{\rm null})$ reveals an important asymmetry: because
$N_{\rm goal}(\theta)$ is maximised at $\theta=0.5$ and decreases as $\theta$ moves
towards $0$ or $1$, the direction of the gap depends on where $\theta_{\rm true}$ lies
relative to $\theta_{\rm null}$.

For $\theta_{\rm null}=0.5$, any deviation of $\theta_{\rm true}$ from $0.5$ implies
$N_{\rm goal}(\theta_{\rm true}) < N_{\rm goal}(\theta_{\rm null})$: the expected stop
iteration of $N_{\rm goal}=599$ is a conservative upper bound on $N_{\rm stop}$,
and experiments stop earlier whenever the truth differs from the null.

The right column of Figure~\ref{fig:stop_iterations_by_truth_conclusive} shows what
happens when the null does not coincide with the precision maximum.
The case $\theta_{\rm null}=0.7$ is representative of a researcher expecting an elevated
base rate (such as a presumed 70\% success probability for an established treatment)
who must judge whether the true rate is practically equivalent to that benchmark.
Unlike the fair coin case, where $\theta_{\rm null}=0.5$ maximises $\theta(1-\theta)$
so any deviation of $\theta_{\rm true}$ reduces the required sample, here the symmetry
breaks: the same deviation can either reduce or increase $N_{\rm goal}$ depending on
its direction.

In particular, when $\theta_{\rm true}<\theta_{\rm null}$ (moving closer to $0.5$),
$N_{\rm goal}(\theta_{\rm true})>N_{\rm goal}(\theta_{\rm null}=0.7)\approx 503$:
the planned budget underestimates the samples required, and
$N_{\rm goal}(\theta_{\rm null})$ acts as a lower bound rather than an upper
one.\footnote{When $\theta_{\rm true}>\theta_{\rm null}=0.7$ (moving further from
$0.5$), $N_{\rm goal}(\theta_{\rm true})<N_{\rm goal}(\theta_{\rm null})$ and
experiments stop sooner than planned.
The PitG median stop iteration visibly exceeds $N_{\rm goal}(\theta_{\rm null})=503$
for $\theta_{\rm true}<0.7$, confirming this underestimate.
DPitG shows qualitatively similar patterns to the $\theta_{\rm null}=0.5$ column
near both ROPE boundaries, elevated above $N_{\rm goal}(\theta_{\rm true})$ where
the conclusiveness gain is largest.}

In general, $N_{\rm goal}(\theta_{\rm null})$ serves as the researcher's advance
planning estimate: a conservative upper bound when $\theta_{\rm null}$ lies at the
precision maximum (e.g.\ $\theta_{\rm null}=0.5$), and a floor otherwise.
The researcher should set $N_{\rm max}$ according to their resources, with
$N_{\rm goal}(\theta_{\rm null})$ as the baseline (here $N_{\rm max}=1{,}500$).

DPitG's $N_{\rm stop}$ equals or exceeds PitG's within the ROPE and in the vicinity
of its boundaries, the extra iterations being precisely what accounts for the higher
conclusiveness seen in Figure~\ref{fig:conclusiveness_rates}. Once $\theta_{\rm true}$
is far enough from the ROPE, DPitG converges to PitG and both stop at
$N_{\rm goal}(\theta_{\rm true})$.

Figure~\ref{fig:stop_conclusiveness_ratios} places the DPitG--PitG trade-off in direct
comparison by overlaying the ratio of conclusiveness rates
(from Figure~\ref{fig:conclusiveness_rates}) and the ratio of $N_{\rm stop}$
(from Figure~\ref{fig:stop_iterations_by_truth_conclusive}) across $\theta_{\rm true}$.\footnote{The
apparent flattening of the stop iteration ratio near ${\rm ROPE}_{\rm max}$ is an artefact of
the finite cap $N_{\rm max}=1{,}500$; with a larger cap both curves would peak sharply
at the boundary.}

The conclusiveness ratio substantially exceeds the $N_{\rm stop}$ ratio throughout,
particularly within and near the ROPE.
Strikingly, the conclusiveness ratio remains above 1 even where the $N_{\rm stop}$
ratio approaches 1 (roughly $0.60 \lesssim \theta_{\rm true} \lesssim 0.63$), confirming
that DPitG gains conclusive outcomes at negligible extra sampling cost in this region.
The lower IQR band confirms that a substantial fraction of DPitG experiments stop
exactly when PitG does.
The fair coin quantitative summary (a $2.6\times$ conclusiveness gain at a median
$4\%$ sampling premium) is consistent with Table~\ref{tab:fair_overall} and
Section~\ref{sec:fair_coin_large_scale}; Table~\ref{tab:goal_impact} shows how the
conclusiveness ratio varies with $\omega_{\rm goal}$.

The DPitG advantage documented here generalises: qualitatively similar conclusiveness
gains hold for $\theta_{\rm null}=0.7$ (as shown in the right panel of
Figure~\ref{fig:stop_iterations_by_truth_conclusive}) and across $\omega_{\rm goal}$
values (see Section~\ref{sec:goal_impact} for the fair coin case).

\section{Discussion}\label{sec:discussion}
DPitG resolves the tension inherent in PitG, requiring both precision and
decisiveness simultaneously, at a modest median sampling cost
(Sections~\ref{sec:fair_coin_large_scale}--\ref{sec:trends}).
We now place this result in methodological context, draw out its practical
implications, and candidly acknowledge its limitations.

DPitG operates under a different philosophy from classical sequential methods.
Group sequential designs \citep{jennison2000} control the familywise Type~I error rate
across interim analyses via alpha-spending functions, and are therefore tied to the
number and timing of planned looks.
DPitG requires no such correction, since the Bayesian posterior accounts for all
accumulated data without reference to the stopping intention.
This property extends to simultaneous tests: the DPitG stopping rule applies
unchanged across independent parallel experiments without multiple-comparisons
correction; Appendix~\ref{app:multiple_comparisons} develops this in full,
together with all three types of Bayesian power analysis (prospective,
retrospective, and replication).

Closest in spirit among Bayesian sequential designs is Bayes Factor Design Analysis
(BFDA; \citealt{schonbrodt2018}), which plans sample size to achieve a target Bayes
Factor with specified probability.\footnote{This is a prospective planning use of the BF, distinct from using the BF as a
sequential stopping rule (see Appendix~\ref{app:bayes_factors} for a discussion
of BF stopping and its limitations).}
The key distinction is that BFDA targets evidential strength (ratio of marginal
likelihoods) whereas DPitG targets posterior precision; the two are complementary
as planning tools, and a design requiring both a conclusive HDI+ROPE framework verdict and a
minimum Bayes Factor is a natural extension.

While comparisons with competing sequential methods clarify DPitG's positioning,
its computational scope deserves equal attention.
DPitG is a generic sequential stopping rule applicable to any posterior admitting a
credible interval, including non-conjugate and hierarchical settings that require MCMC;
the present analysis restricts to conjugate posteriors, where the HDI is fully
analytical and no specialist probabilistic programming software is required.

A recurring concern about PitG (and by extension DPitG) is that the precision
requirement can seem too demanding in practice.
Researchers frequently underestimate the noise in their measurements and are consequently surprised by the sample sizes that
$\omega_{\rm goal}$ implies (J.K.\ Kruschke, pers.\ comm.\ 2022). This is not a flaw
in the method but a form of epistemic honesty: imprecise posteriors produce inconclusive
results regardless of the algorithm applied; DPitG simply refuses to declare a verdict
until the posterior is genuinely informative. Setting $\omega_{\rm goal}$ thus forces
realistic planning rather than masking insufficient power behind a false reject (i.e., false positive) or a
silently inconclusive stop.
This discipline is a feature, not a limitation, for any analyst whose conclusions
must rest on genuinely informative data.

DPitG serves researchers in two complementary roles.
As an operational stopping rule, it is best suited to settings where data
collection is relatively cheap and the budget comfortably exceeds $N_{\rm goal}$:
large-scale digital A/B testing \citep{johari2022}, opinion polling, and automated
quality-control pipelines are typical examples.
As a planning tool, it provides a diagnostic function even in resource-constrained
settings such as longitudinal studies, clinical trials, or expensive policy evaluations:
$N_{\rm goal}$ makes explicit what precision is achievable within the available budget,
enabling an honest assessment of what conclusions the data can and cannot support,
both prospectively (before collection begins) and retrospectively (to audit the
precision of a completed study).

The ROPE must be grounded in domain knowledge rather than chosen to suit the data
\citep{kruschke2018}; Appendix~\ref{app:practical_guidance} discusses principled
approaches for setting it, including the MCID, bioequivalence standards, and
sensitivity analyses, together with three broader considerations supporting
DPitG's adoption (equivalence testing, pre-registration, and the replication crisis).
Together with $\omega_{\rm goal}$ and $N_{\rm max}$, the ROPE completes the three
design parameters that must be fixed before data collection begins.

With those design inputs committed, the joint operating characteristics of DPitG can
be assessed prospectively: a researcher can check, analytically or via simulation,
that the planned $N_{\rm max}$ achieves the target conclusiveness rate at the effect
sizes of interest.
As the Results demonstrate, error rates are negligible except near the ROPE boundaries,
where neither an accept nor a reject verdict is easily justified.
A key practical implication (Section~\ref{sec:goal_impact}) is that conclusiveness
is more sensitive to $N_{\rm max}$ than to $\omega_{\rm goal}$: for the fair coin
case ($\Delta_{\rm ROPE}=0.1$), the median sampling premium relative to $N_{\rm goal}$
is zero for $\omega_{\rm goal}\le 0.07$ and rises to 41\% at $\omega_{\rm goal}=0.10$,
so setting $N_{\rm max}$ generously is the primary practical lever for ensuring high
conclusiveness across a range of precision goals.

We now turn to the limits of the present work.
Four limitations deserve acknowledgement.
First, the results are sensitive to the three design parameters (the ROPE,
$\omega_{\rm goal}$, and $N_{\rm max}$), all of which must be committed to before
data collection begins.
A narrower ROPE (and with it a tighter $\omega_{\rm goal}$) increases the sample size
required for decisiveness and raises the
inconclusive rate near its boundaries.
Second, the stopping properties are validated empirically only for binary (Bernoulli)
data under a conjugate Beta prior.
The normal approximation underpinning $N_{\rm goal}$
(Equation~\ref{eq:pitg_stop_iteration}) may underestimate the required sample size
in small-sample settings or for highly skewed likelihoods; in such cases the exact
Binomial result in Appendix~\ref{app:expected_stop} should be preferred.
Third, the present analysis uses a flat (uniform) prior throughout, as a conservative
choice that avoids imposing unverified assumptions.
The flat prior contributes no pseudo-observations, requiring more data to achieve
$\omega_{\rm goal}$ than a well-justified informative prior would \citep[][Ch.~2]{gelman2013bayesian}.
Informative priors, where available and well-justified, can therefore serve as a
practical mitigation by narrowing the starting HDI width and lowering the required
sample size.
Conversely, a materially incorrect informative prior may bias the posterior in a way
that the precision criterion alone cannot detect \citep[][Ch.~6]{gelman2013bayesian};
we defer a full treatment of principled prior selection to a later study.
Fourth, while DPitG applies in principle to any posterior admitting such an interval,
this paper does not characterise its behaviour when the HDI is estimated from MCMC
samples.
In such settings the HDI estimate carries Monte Carlo error \citep{kruschke2015doing}
that could affect stopping
calibration, particularly in early data collection when the posterior is diffuse,
and this remains an open question for subsequent investigation.

\section{Conclusion}\label{sec:conclusion}

We propose DPitG as a sequential stopping rule that jointly requires posterior
precision and a conclusive verdict before data collection ends.
PitG eliminates the early-peeking bias of coupled methods by decoupling the stopping
and decision criteria \citep{kruschke2015doing}, but accepts high inconclusive rates
as the price. Inconclusive rates rise steeply the closer the null hypothesis is to the true value.

DPitG closes this gap by imposing both requirements simultaneously: data collection
continues until $\omega_{\rm HDI} \le \omega_{\rm goal}$ \textit{and} the posterior's
position relative to the ROPE yields a conclusive verdict.
This joint requirement delivers a $2.6\times$ gain in conclusiveness over PitG at a
median cost of only 4.7\% (4\% among conclusive experiments) more samples
($\theta_{\rm true}=0.5$, $\omega_{\rm goal}=0.08$,
${\rm ROPE}_{\rm min}^{\rm max}=[0.45,0.55]$),
while the HDI+ROPE algorithm accrues a systematic false-positive rate because its
coupled design can stop on wide, early posteriors before they are representative.
The gain is robust: DPitG's conclusiveness holds nearly flat ($97.3\%$--$98.0\%$)
across the same $\omega_{\rm goal}$ range, over which PitG's conclusiveness
varies widely\footnote{And collapses to zero when $\omega_{\rm goal} = {\rm ROPE}_{\rm max} - {\rm ROPE}_{\rm min}$.}
and is largest near the ROPE boundaries where evidence is hardest to read.

Unlike NHST, which can only reject or fail to reject the null (see Appendix~\ref{app:nhst}),
the HDI+ROPE framework's decision rule enables positive \textit{acceptance} of the null
when the entire HDI falls within the ROPE, a critical advantage for confirmatory
research areas such as bioequivalence, safety testing, and software regression testing;
DPitG reliably delivers this verdict.

Prospective planning is straightforward: the closed-form
$N_{\rm goal}(\theta_{\rm null}, \omega_{\rm goal}) \propto
\theta_{\rm null}(1-\theta_{\rm null})/\omega_{\rm goal}^2$
provides the minimum budget required, with $N_{\rm max}$ set generously above this
baseline to ensure DPitG reaches its full conclusiveness potential.

The framework extends to single-group continuous data
(Appendix~\ref{app:expected_stop}) and to two-group comparisons, both binomial
and continuous (Appendix~\ref{app:extensions}).\footnote{These extensions are
theoretical derivations; stopping-property validation beyond the single-group Bernoulli
case remains to be completed in a dedicated simulation study.}
All algorithms are implemented in openly available
code\footnote{\href{https://github.com/elzurdo/dpitg}{https://github.com/elzurdo/dpitg}},
with an interactive online calculator for prospective sample-size planning and
retrospective analysis available at
\href{https://r-w-t-y.streamlit.app}{https://r-w-t-y.streamlit.app}.

DPitG is the method of choice whenever a reliable verdict (acceptance, rejection,
or inconclusive) is required and the budget is sufficient to reach
$N_{\rm goal}(\theta_{\rm null})$; the closed-form $N_{\rm goal}$ makes this judgement
transparent before any data are collected.

\appendix

\section{HDI and ETI: Credible Interval Variants}\label{app:hdi}

Given a posterior density $p(\theta \mid \mathbf{x})$, where $\theta$ is the parameter
of interest and $\mathbf{x}$ the observed data, a $(1-\alpha)$ credible interval
$[L, U]$ satisfies
\begin{equation}\label{eq:credible_interval}
  \int_L^U p(\theta \mid \mathbf{x})\,d\theta = 1 - \alpha.
\end{equation}
Infinitely many intervals satisfy this constraint. This paper uses the Highest Density
Interval throughout; the Equal-Tailed Interval is described below for comparison.

\subsection*{Highest Density Interval (HDI)}

The HDI is the shortest interval satisfying Equation~\ref{eq:credible_interval}:
\begin{equation}\label{eq:hdi}
  [L^*, U^*] = \underset{[L,U]:\;\int_L^U p(\theta|\mathbf{x})\,d\theta\,=\,1-\alpha}{\arg\min}\;(U - L).
\end{equation}
An equivalent characterisation for unimodal posteriors: $[L^*, U^*]$ is the unique
interval for which $p(L^* \mid \mathbf{x}) = p(U^* \mid \mathbf{x})$, so every point
inside has posterior density at least as high as every point outside (hence ``highest
density'').\footnote{For multimodal posteriors, multiple intervals can satisfy the
equal-density boundary condition, and the HDI may be a union of disjoint intervals;
the shortest-interval definition in Equation~\ref{eq:hdi} selects the one with minimum
total width. All posteriors in this paper are unimodal.}
For the Binomial model used in this work, under a flat prior $\mathrm{Beta}(1,1)$, the
posterior after $s$ successes and $f$ failures is $\mathrm{Beta}(s+1,\,f+1)$, and the
HDI is found by minimising the interval width over the lower tail
probability.\footnote{Specifically, minimise
  $F^{-1}(1-\alpha+\ell)-F^{-1}(\ell)$ over $\ell\geq 0$; the minimiser $\ell^*$
  gives $L^*=F^{-1}(\ell^*)$ and $U^*=F^{-1}(1-\alpha+\ell^*)$
  (\texttt{HDIofICDF} in the accompanying code).}

\subsection*{Equal-Tailed Interval (ETI)}

The ETI places $\alpha/2$ probability mass in each tail:
\begin{equation}\label{eq:eti}
  L_{\rm ETI} = F^{-1}\!\left(\tfrac{\alpha}{2}\right), \qquad
  U_{\rm ETI} = F^{-1}\!\left(1 - \tfrac{\alpha}{2}\right),
\end{equation}
where $F^{-1}$ is the posterior quantile function. The ETI is simple to compute and
uniquely defined, but it is not generally the shortest interval.
For symmetric, unimodal posteriors the HDI and ETI coincide; they diverge for skewed
distributions (for instance, a Beta posterior near $\theta = 0$ or $\theta = 1$)
where the HDI is noticeably shorter and therefore preferred.

\section{Regarding NHST}\label{app:nhst}
Null hypothesis significance testing (NHST, \citealt{fisher1925, neymanpearson1933})
is included here as contextual background rather than as a competing method.
A p-value measures the probability of observing data at least as extreme as those
obtained \emph{assuming the null is true}, $P(\mathrm{data}\mid H_0)$, not the
probability that the null is true given the data, $P(H_0\mid\mathrm{data})$
\citep{cohen1994, wasserstein2016};
consequently, NHST can only reject the null or remain inconclusive; it carries
no mechanism to positively accept it, making it structurally unsuitable for the
precision-and-decisiveness criterion that motivates DPitG.
Understanding its sequential behaviour nevertheless clarifies why the Bayesian
HDI+ROPE framework was developed.

In NHST, at each iteration the experiment stops and $\theta_{\rm null}$ is rejected
whenever the p-value falls below the false positive rate threshold $\alpha_{\rm FPR}$.
Since the decision rule is based solely on this single threshold, data that fail to
trigger rejection yield an inconclusive result rather than evidence in favour of the null.

Using the same fair coin setup ($\theta_{\rm true}=\theta_{\rm null}=0.5$, $N_{\rm min}=30$) described
in the Methods section (\ref{sec:methods}), we apply NHST with $\alpha_{\rm FPR}=0.05$
as the coupled stopping and decision criterion. A reliable sequential method should maintain a
low rejection rate throughout when the null hypothesis is true: as demonstrated in
Section~\ref{sec:fair_coin_large_scale}, the HDI+ROPE algorithm's rejection rate quickly asymptotes near 6\% and
both precision-based methods maintain a zero rejection rate.

\begin{figure}[h!]
  \centering
  \includegraphics[width=1\textwidth]{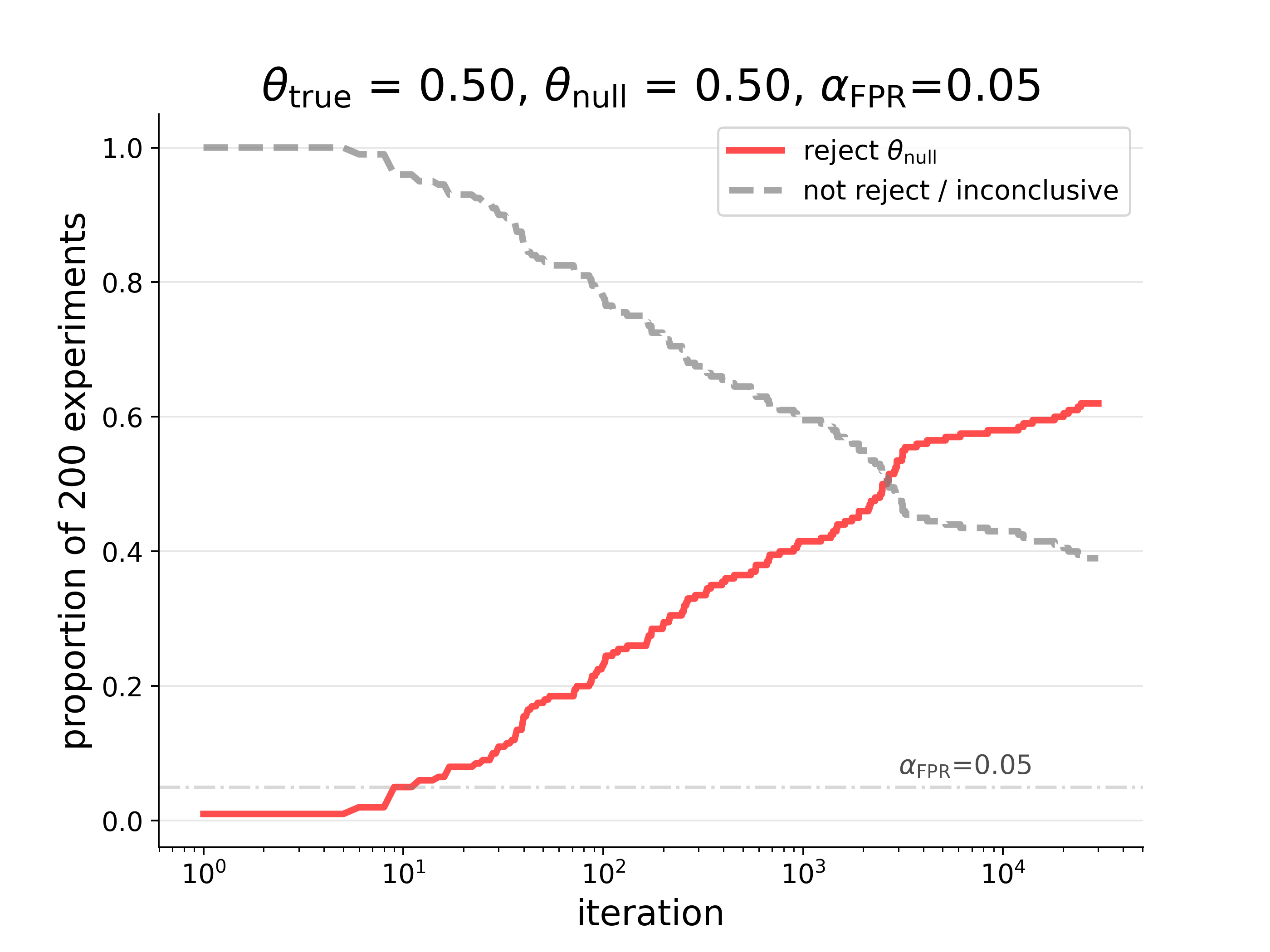}
  \caption{Cumulative decision rates across iterations for $M=200$ fair coin experiments
  ($\theta_{\rm true}=\theta_{\rm null}=0.5$, $\alpha_{\rm FPR}=0.05$,
  $N_{\rm max}=30{,}000$), extending Figure~\ref{fig:fair_decisions} with NHST
  as a fourth panel; it appears in this appendix as NHST lies outside the scope
  of the main analysis.
  At each iteration $N$, the plotted proportions count all experiments that reached
  a decision at or before $N$; their sum is 100\% at every point.
  \textit{Red solid line}: reject $\theta_{\rm null}$.
  \textit{Gray dashed line}: inconclusive.
  The extended horizon ($N_{\rm max}=30{,}000$ vs.\ $1{,}500$ in the main analysis)
  is needed to make the growing false-rejection trend visible; the reduced experiment
  count ($M=200$ vs.\ $5{,}000$) reflects the higher computational cost of evaluating
  p-values relative to HDI-based criteria, which also accounts for the stepped
  appearance of the curves.}
  \label{fig:fair_decisions_nhst}
\end{figure}

Figure~\ref{fig:fair_decisions_nhst} shows the opposite for NHST: the rejection rate
(red solid) grows steadily over time, while the inconclusive proportion (gray dashed)
decreases. To make this trend visible, the simulation is extended to
$N_{\rm max}=30{,}000$ iterations, well beyond the $N_{\rm max}=1{,}500$ used in
the main analysis, since false rejections accumulate slowly and the growing trend
would not be apparent at shorter horizons.

As pointed out by \citet{kruschke2015doing}, as evidence accumulates under NHST
the probability of observing a sample extreme enough to trigger rejection grows without
bound, even when the null hypothesis is true. To paraphrase: ``with infinite
peeking, the false positive rate approaches 100\%.'' This is because the stopping
rule is tied directly to the decision rule with no reference to an effect size: the
more data collected, the more likely an extreme sample will appear. Continuously
monitoring the data and applying a fixed p-value threshold therefore leads to an
inflated false-positive (Type~I) error rate, a manifestation of the multiple-testing
problem
\citep{simmons2011}.
Lowering $\alpha_{\rm FPR}$ delays but does not prevent this accumulation: for any
fixed threshold, the false rejection rate still grows without bound as data continue
to be collected.

\section{Derivation of Expected Precision Goal Stop Iteration}\label{app:expected_stop}

The precision-based stopping algorithms (PitG \citealt{kruschke2015doing} and DPitG) halt when the HDI width
(see Appendix~\ref{app:hdi})
falls below the precision goal $\omega_{\rm goal}$. Because the posterior HDI width
shrinks as $N^{-1/2}$, one can predict in advance the iteration at which this
threshold will be crossed. The General CLT Form below uses a normal approximation
and is intended as an analytical guide; the Binomial Case refines this using the
exact Beta posterior variance, and the Continuous Case provides two variants
(CLT and Student-$t$).

\subsection*{General CLT Form}

Assume an estimator $\hat\theta_N$ with asymptotic distribution
\begin{equation}\label{eq:clt_general}
\hat\theta_N \approx \mathcal{N}\!\left(\theta,\,\frac{V(\theta)}{N}\right),
\end{equation}
where $V(\theta)$ is the per-observation variance contribution. For a symmetric
central interval with critical value $z_{*}$ (e.g., $z_{*}=1.96$ for 95\%), the
interval width is approximately
\begin{equation}\label{eq:hdi_width_approx}
w \approx 2 z_{*}\sqrt{\frac{V(\theta)}{N}}.
\end{equation}
Setting $w = \omega_{\rm goal}$ and solving for $N$ gives the general expected stop
iteration:
\begin{equation}\label{eq:expected_stop_iteration_clt}
N_{\rm goal} \approx \frac{4 z_{*}^{2}}{\omega_{\rm goal}^{2}}\,V(\theta).
\end{equation}

\subsection*{Binomial Case}

For Bernoulli outcomes the posterior under the Haldane prior $\mathrm{Beta}(0,0)$
is $\mathrm{Beta}(s,\,f)$, where $s$ is the number of successes and
$f = N - s$ the number of failures.\footnote{This appendix uses the
Haldane prior rather than the flat $\mathrm{Beta}(1,1)$ prior of
Appendix~\ref{app:hdi}: the Haldane convention yields the clean $-1$ correction
in the planning formula and is consistent with the planning function in the
accompanying code (\texttt{binomial\_rate\_ci\_width\_to\_sample\_size} in
\texttt{utils\_stats.py}).}
Its exact variance is
\[
\mathrm{Var}(\theta \mid s, f) = \frac{\hat\theta\,(1-\hat\theta)}{N+1},
\]
where $\hat\theta = s/N$. Substituting into Equation~\ref{eq:hdi_width_approx}
and setting $w = \omega_{\rm goal}$ gives the exact expected stop iteration:
\begin{equation}\label{eq:expected_stop_iteration_binomial}
N_{\rm goal} = \frac{4 z_{*}^{2}}{\omega_{\rm goal}^{2}}\,\theta(1-\theta) - 1.
\end{equation}
The $-1$ arises from the $N+1$ denominator in the exact Beta posterior variance,
whereas the CLT approximation (Equation~\ref{eq:expected_stop_iteration_clt})
treats the denominator as $N$ and omits it.
As shown in Figure~\ref{fig:min_sample_by_goal} in Section~\ref{sec:expected_stop_iteration},
$N_{\rm goal}$ is maximised at $\theta=0.5$, confirming that the fair coin case requires the
largest sample to achieve a given precision goal, and decreases monotonically as
$\theta$ moves towards $0$ or $1$.
Equation~\ref{eq:pitg_stop_iteration} in the main text uses the CLT approximation
(Equation~\ref{eq:expected_stop_iteration_clt}), which replaces $N+1$ with $N$ and
drops the $-1$ term, negligible for large $N$.

\subsection*{Continuous Case}

For continuous outcomes when the estimand is a mean $\mu$ with variance
$\sigma^{2}$, substituting $V(\mu)=\sigma^{2}$ (estimated by sample variance
$s^{2}$) into Equation~\ref{eq:expected_stop_iteration_clt} gives the
CLT-based expected stop iteration:
\begin{equation}\label{eq:expected_stop_iteration_mean}
N_{\rm goal} \approx \frac{4 z_{*}^{2}\sigma^{2}}{\omega_{\rm goal}^{2}}.
\end{equation}
This approximation is accurate whenever $N \ge 30$, where the Student-$t$
critical value approaches $z_{*}$.\footnote{
  For small samples ($N < 30$), the exact posterior for $\mu$ under a
  non-informative prior is a scaled Student-$t$ with $\nu = N-1$ degrees
  of freedom, giving HDI width $w = 2\,t_{\alpha/2,\,N-1}\,s/\!\sqrt{N}$,
  where $\alpha = 1 - \text{CI fraction}$ (e.g.\ $\alpha = 0.05$ for
  95\%).  Setting $w = \omega_{\rm goal}$ yields an implicit equation in
  $N$ that must be solved numerically.  Because precision goals are seldom
  met at such small $N$, this refinement is not recommended for
  sample-size planning.}

\section{Precision-Based Sample Size Planning for Two-Group Settings}\label{app:extensions}

The HDI+ROPE framework and precision-goal approach presented in the main text for single-group
Bernoulli data generalises to two-group comparisons.
This appendix outlines the key adaptations for two-group binomial and continuous
outcomes, together with the corresponding sample-size planning equations.
For single-group continuous data, the expected stop iteration is derived in
Appendix~\ref{app:expected_stop}.
Empirical validation of DPitG in these two-group settings is left for future work.

\subsection*{Two-Group Comparisons}

For two-group comparisons, whether binomial ($\theta_1 - \theta_2$) or continuous
($\mu_1 - \mu_2$), the standard error of the difference serves as the precision
metric, enabling sequential two-group testing under the same HDI+ROPE decision framework
\citep{kruschke2013}.
For the binomial case with conjugate Beta posteriors, the posterior of the difference
can be evaluated by direct i.i.d.\ sampling from each Beta distribution separately,
without Markov Chain Monte Carlo (MCMC; \citealt{gelman2013bayesian}).
For the continuous case, under a weakly informative prior the posterior of
$\mu_1 - \mu_2$ converges to the frequentist sampling distribution by the
Bernstein--von Mises theorem \citep{gelman2013bayesian}, so the standard error from
Welch's approximation provides a reliable surrogate for the posterior HDI width when
$N \gtrsim 30$ (see Appendix~\ref{app:expected_stop}).

\subsubsection*{Two-Group Binomial}

With independent Beta posteriors for $\theta_1$ and $\theta_2$, the HDI width of
the difference $\theta_1 - \theta_2$ is approximated via the standard error
\begin{equation}\label{eq:se_diff_binomial}
    \text{SE}(\hat\theta_1 - \hat\theta_2) \approx
    \sqrt{\frac{\theta_1(1-\theta_1)}{N_1} + \frac{\theta_2(1-\theta_2)}{N_2}}.
\end{equation}
Here $\theta_i$ is replaced in practice by the posterior mean $\hat\theta_i = s_i/N_i$;
the exact Beta posterior variance replaces $N_i$ with $N_i+1$ (cf.\ the Binomial Case
in Appendix~\ref{app:expected_stop}).
Setting the HDI width $2z_*\,\text{SE} = \omega_{\rm goal}$ (where $z_*$ is the
normal critical value for the chosen credible interval level;
see Appendix~\ref{app:expected_stop}) and writing
$N_1 = rN_2$ for group-size ratio $r = N_1/N_2 > 0$, solving for $N_2$ gives
\begin{equation}\label{eq:ngoal_two_group_binomial}
    N_2 \approx \frac{4 z_{*}^{2}}{\omega_{\rm goal}^{2}}
        \left(\frac{\theta_1(1-\theta_1)}{r} + \theta_2(1-\theta_2)\right),
    \quad N_1 = r N_2.
\end{equation}
For equal group sizes ($r=1$, $N_1=N_2=N$) this simplifies to
\begin{equation}\label{eq:ngoal_two_group_binomial_equal}
    N \approx \frac{4 z_{*}^{2}[\theta_1(1-\theta_1)+\theta_2(1-\theta_2)]}{\omega_{\rm goal}^{2}}.
\end{equation}

\subsubsection*{Two-Group Continuous}

Following Welch's approximation, the standard error of the difference of means is
\begin{equation}\label{eq:se_diff_continuous}
    \text{SE}(\hat\mu_1 - \hat\mu_2) \approx
    \sqrt{\frac{\sigma_1^2}{N_1} + \frac{\sigma_2^2}{N_2}}.
\end{equation}
Writing $N_1 = rN_2$ for group-size ratio $r = N_1/N_2 > 0$ and solving for $N_2$ gives
\begin{equation}\label{eq:ngoal_two_group_continuous}
    N_2 \approx \frac{4 z_{*}^{2}}{\omega_{\rm goal}^{2}}
        \left(\frac{\sigma_1^2}{r} + \sigma_2^2\right),
    \quad N_1 = r N_2,
\end{equation}
with $\sigma_i^2$ replaced by $s_i^2$ when the population variances are unknown.
For equal group sizes ($r=1$, $N_1=N_2=N$) this simplifies to
\begin{equation}\label{eq:ngoal_two_group_continuous_equal}
    N \approx \frac{4 z_{*}^{2}(\sigma_1^2+\sigma_2^2)}{\omega_{\rm goal}^{2}}.
\end{equation}

\section{Multiple Comparisons and Bayesian Power Analysis}\label{app:multiple_comparisons}

\subsection*{Multiple Comparisons}

A practical advantage of the HDI+ROPE decision framework over frequentist approaches concerns
multiple comparisons. In NHST, conducting several tests simultaneously inflates the
false-positive error rate (Type~I), requiring corrections (e.g.\ Bonferroni) that reduce the power of
each individual test. Frequentist confidence intervals used for simultaneous inference inherit the same
dependence on the full set of intended comparisons, since their simultaneous coverage
properties are tied to p-values. Consequently, both power and precision planning must account for the number
of planned tests.

The Bayesian posterior is not affected in this way. Because the posterior distribution
is determined solely by the observed data and the assumed data-generating process, it
is unaffected by how many other comparisons were planned, unlike frequentist
p-values, which depend on the full set of intended tests even when some are never
conducted \citep{gelman2012, kruschke2015doing}.
The power of any one test is therefore unaffected by how many other comparisons are
planned, provided each test uses an independent model.
An independent model means each comparison is fitted separately, without shared
parameters or a common prior; when tests are connected through a hierarchical model,
posteriors are linked across comparisons and the independence no longer holds
\citep{gelman2012}.
The precision goal $\omega_{\rm goal}$ and the stopping rule of DPitG therefore apply
unchanged in a multiple-testing context, without the need for correction.

\subsection*{Three Types of Bayesian Power Analysis}

The same design flexibility extends to how power is conceptualised in the Bayesian
setting, where \citet{kruschke2015doing} distinguishes three types of power analysis that arise in
the Bayesian setting.

\textit{Prospective} power analysis is conducted before data collection: given a
hypothesised data-generating distribution (from theory, pilot data, or prior research),
one simulates experiments and asks how often the precision goal $\omega_{\rm goal}$ is
achieved within the budget $N_{\rm max}$.
The analytical formula for $N_{\rm goal}$ (Equation~\ref{eq:pitg_stop_iteration})
provides this as a closed-form lower bound on the required sample size; DPitG may
require additional samples when the true parameter value lies near a ROPE boundary
(see Section~\ref{sec:trends}).

\textit{Retrospective} power analysis uses the posterior from already-collected data as
the hypothetical data generator to ask whether the completed study was sufficiently
powered. In NHST this quantity is a monotone function of the p-value and therefore adds
no new information beyond it \citep{hoenig2001}; in the Bayesian setting it can reveal
whether the precision goal $\omega_{\rm goal}$ was realistically achievable given the
observed data, independently of the decisiveness verdict.

\textit{Replication} power asks: if the experiment were repeated exactly, what is the
probability of achieving the goal in the new study, informed by the results of the
first? In the DPitG framework this is straightforward to compute by using the initial
posterior as the prior for the follow-up and re-running the stopping rule; NHST has
no principled access to a posterior-derived data generator and cannot support this
form of analysis \citep{kruschke2015doing}.

\section{Regarding Bayes Factors}\label{app:bayes_factors}

This appendix concerns the use of the Bayes Factor (BF) as a \textit{sequential
stopping rule}; this role is distinct from its use in Bayes Factor Design Analysis
(BFDA), which is a prospective planning tool discussed in Section~\ref{sec:discussion}.

The BF is a widely used Bayesian measure for comparing two hypotheses.
For a null hypothesis $H_0$ and an alternative $H_1$, the Bayes Factor in favour of
the null is defined as the ratio of marginal likelihoods:
\begin{equation}\label{eq:bf_definition}
  \mathrm{BF}_{01} = \frac{p(\mathrm{data} \mid H_0)}{p(\mathrm{data} \mid H_1)}.
\end{equation}
A value $\mathrm{BF}_{01} > 1$ constitutes evidence in favour of $H_0$; a value
$\mathrm{BF}_{01} < 1$ constitutes evidence against it.
The scale proposed by \citet{jeffreys1961} and refined by \citet{kassraftery1995}
provides conventional thresholds: $\mathrm{BF}_{01} > 3$ is typically taken as
``substantial'' evidence for the null, and $\mathrm{BF}_{01} < 1/3$ as substantial
evidence against it.

For the Bernoulli setting with $z$ successes in $N$ trials, \citet{kruschke2015doing}
(Chapter~13.3.2) uses a point null $H_0\colon \theta = \theta_{\rm null}$ and a uniform
alternative $H_1\colon \theta \sim \mathrm{Beta}(1, 1)$.  The marginal likelihood under
$H_0$ is simply $\theta_{\rm null}^{z}(1-\theta_{\rm null})^{N-z}$, and under $H_1$ it
is the Beta--Binomial integral
\begin{equation}\label{eq:bf_h1_marginal}
  p(\mathrm{data} \mid H_1)
  = \int_0^1 \theta^{z}(1-\theta)^{N-z}\,d\theta
  = B(z+1,\, N-z+1),
\end{equation}
where $B(\cdot,\cdot)$ denotes the Beta function, giving the closed-form expression
\begin{equation}\label{eq:bf_bernoulli}
  \mathrm{BF}_{01}
  = \frac{\theta_{\rm null}^{z}(1-\theta_{\rm null})^{N-z}}{B(z+1,\, N-z+1)}.
\end{equation}
The sequential stopping rule is: continue sampling until either
$\mathrm{BF}_{01} \geq 3$ (accept $H_0$) or $\mathrm{BF}_{01} \leq 1/3$ (reject $H_0$),
or the budget $N_{\rm max}$ is exhausted.

\subsection*{Sequential Behaviour}

Unlike frequentist $p$-values, BF stopping is valid under optional stopping: because
the Bayes Factor depends only on the observed data and the assumed model, not on the
sampling intention or the number of interim looks; sequential monitoring does not
inflate it \citep{rouder2014}.
The question is therefore not whether BF stopping is theoretically principled, but
whether it achieves adequate operating characteristics in practice.

\citet{kruschke2015doing} evaluates those characteristics across $M=1{,}000$ simulated
sequences of up to $N_{\rm max}=1{,}500$ flips, comparing BF stopping against NHST,
the HDI+ROPE algorithm, and PitG (his Figures~13.6 and~13.7).
The main findings relevant to this paper are as follows.

\subsubsection*{When the null is true ($\theta_{\rm true}=\theta_{\rm null}=0.5$)}

This is the setting shown in Kruschke's Figure~13.6 (for which
Figure~\ref{fig:fair_decisions} of this paper is analogous).
The BF stopping rule converges to an asymptotic false rejection rate
(i.e, the \textit{false alarm rate})
of just over 20\%, lower than NHST, which diverges to 100\%, but substantially higher
than PitG, which maintains a 0\% false rejection rate \citep{kruschke2015doing}.
As demonstrated in this paper, DPitG also achieves a 0\% false rejection rate.
The BF correctly accepts the null for the remaining sequences, although it can do
so on the basis of precise or imprecise data indiscriminately.

\subsubsection*{When the null is false ($\theta_{\rm true}=0.65, \ \theta_{\rm null}=0.5$)}

This setting is displayed in Kruschke's Figure~13.7 (which is captured in the middle
panel of Figure~\ref{fig:conclusiveness_rates}).
The BF falsely accepts the null in approximately 40\% of sequences despite the true
parameter lying outside the ROPE, an alarmingly higher false acceptance rate,
whereas the precision-based methods yield 0 false acceptances.

\subsubsection*{Parameter estimation at the stopping point}

Because the BF stopping rule fires on the \emph{extremeness} of the likelihood ratio
rather than the \emph{width} of the posterior, it can accept or reject with very little
data.  As the three lower rows of Figures 13.6 and 13.7 of \citet{kruschke2015doing} illustrate, $\hat{\theta}$ at stopping is noticeably
biased away from $\theta_{\rm true}$ in both the accept and the reject case, more so
than the HDI+ROPE algorithm, and substantially more than PitG.  This is the fundamental problem
identified in \citet{kruschke2015doing}: a stopping rule based on extremeness in the data
automatically biases the sample toward extreme estimates; only a precision-based stopping
rule guarantees near-unbiased parameter estimation.

\subsection*{Key Limitations in Context}

Three structural limitations follow from the BF's definition.

\subsubsection*{No effect-size criterion}

$\mathrm{BF}_{01}$ compares the point null $\theta = \theta_{\rm null}$ against the
prior-averaged alternative.  While the alternative prior implicitly encodes a distribution
over effect sizes, the BF provides no explicit mechanism to require the posterior to fall
cleanly inside or outside a practically meaningful interval: a parameter value just
outside the ROPE is treated identically to one that is far from it.  The HDI+ROPE
decision framework addresses this explicitly by requiring the HDI to fall cleanly inside
or outside the ROPE before a verdict is issued.

\subsubsection*{Sensitivity to the alternative prior}

$\mathrm{BF}_{01}$ depends on the choice of prior $p(\theta \mid H_1)$
(Equation~\ref{eq:bf_h1_marginal}).  Kruschke's choice of a uniform
$\mathrm{Beta}(1,1)$ prior is one convention; different priors yield different
numerical BF values and therefore different stopping points, even for identical data.
The HDI+ROPE stopping rule has no such ambiguity: its behaviour depends only on the
posterior and the pre-specified ROPE and $\omega_{\rm goal}$.

\subsubsection*{No precision guarantee}

The BF can stop early on imprecise posteriors.  In contrast, PitG and DPitG require
the HDI width to fall below $\omega_{\rm goal}$ before any verdict is issued, providing
a structural guarantee that accepted or rejected conclusions rest on adequately precise
estimates.

These limitations notwithstanding, BF stopping and HDI+ROPE are not mutually exclusive.
As noted in Section~\ref{sec:discussion}, a design requiring both a conclusive
HDI+ROPE verdict \textit{and} a minimum Bayes Factor is a natural extension of DPitG,
combining the precision guarantee of the latter with the evidential strength measure
of the former.

\section{Practical Design Guidance and Broader Context}\label{app:practical_guidance}

\subsection*{Setting the ROPE}

Three design inputs must be fixed before data collection: the ROPE, the precision goal
$\omega_{\rm goal}$, and the maximum sample size $N_{\rm max}$.
The precision goal should satisfy $\omega_{\rm goal} \leq \Delta_{\rm ROPE}$ and directly
determines the expected stop iteration $N_{\rm goal}$
(Equation~\ref{eq:pitg_stop_iteration}); $N_{\rm max}$ is set by budget.
Of these three subjective inputs the ROPE is the most consequential.
It should be determined from domain knowledge rather than from the data,
a requirement that forces the researcher to commit to a practically meaningful
effect-size threshold before observing outcomes \citep{kruschke2018}.
In clinical research the Minimal Clinically Important Difference (MCID) provides a
principled basis \citep{jaeschke1989}; in bioequivalence testing, regulatory guidelines define the equivalence
margin \citep{schuirmann1987}; in industrial quality
control, engineering tolerances provide an equivalent principled basis.
When no domain standard exists, sensitivity analyses across a range of ROPE widths
are recommended before committing to a single value \citep{lakens2018}.

\subsection*{Broader Adoption Considerations}

Three considerations support broader adoption of the DPitG framework.

First, formally accepting the null via the HDI+ROPE framework's criterion parallels the objective
of frequentist equivalence testing through the Two One-Sided Tests (TOST) procedure
\citep{schuirmann1987, lakens2018}, providing a Bayesian counterpart to a methodology
established in regulatory and applied settings \citep{fda2010bayesian}.

Second, because $\omega_{\rm goal}$, the ROPE, and $N_{\rm max}$ must all be specified
before data collection begins, DPitG is structurally compatible with pre-registration
standards \citep{nosek2018}, directly reducing the researcher degrees of freedom,
the analytic choices that enable post-hoc optimisation of results, that drive
selective reporting.

Third, the replication crisis has been partly attributed to underpowered studies whose
imprecise estimates fail to replicate \citep{osc2015, gelman2014}; DPitG addresses this
at the design stage by making precision a non-negotiable stopping criterion rather than
an aspiration, one that is structurally enforced by construction.

\bibliography{references.bib}

\begin{thebibliography}{}

\bibitem[Cohen, 1988]{cohen1988}
Cohen, J. (1988).
\newblock {\em Statistical Power Analysis for the Behavioral Sciences}.
\newblock Lawrence Erlbaum Associates, Hillsdale, NJ, 2nd edition.

\bibitem[Cohen, 1994]{cohen1994}
Cohen, J. (1994).
\newblock The earth is round ($p < .05$).
\newblock {\em American Psychologist}, 49(12):997--1003.

\bibitem[Fisher, 1925]{fisher1925}
Fisher, R.~A. (1925).
\newblock {\em Statistical Methods for Research Workers}.
\newblock Oliver and Boyd, Edinburgh.

\bibitem[Gelman and Carlin, 2014]{gelman2014}
Gelman, A. and Carlin, J. (2014).
\newblock Beyond power calculations: Assessing {Type S} (sign) and {Type M} (magnitude) errors.
\newblock {\em Perspectives on Psychological Science}, 9(6):641--651.

\bibitem[Gelman et~al., 2013]{gelman2013bayesian}
Gelman, A., Carlin, J.~B., Stern, H.~S., Dunson, D.~B., Vehtari, A., and Rubin, D.~B. (2013).
\newblock {\em Bayesian Data Analysis}.
\newblock Chapman and Hall/{CRC}, Boca Raton, FL, 3rd edition.

\bibitem[Gelman et~al., 2012]{gelman2012}
Gelman, A., Hill, J., and Yajima, M. (2012).
\newblock Why we (usually) don't have to worry about multiple comparisons.
\newblock {\em Journal of Research on Educational Effectiveness}, 5(2):189--211.

\bibitem[Hoenig and Heisey, 2001]{hoenig2001}
Hoenig, J.~M. and Heisey, D.~M. (2001).
\newblock The abuse of power: The pervasive fallacy of power calculations for data analysis.
\newblock {\em The American Statistician}, 55(1):19--24.

\bibitem[Jaeschke et~al., 1989]{jaeschke1989}
Jaeschke, R., Singer, J., and Guyatt, G.~H. (1989).
\newblock Measurement of health status: Ascertaining the minimal clinically important difference.
\newblock {\em Controlled Clinical Trials}, 10(4):407--415.

\bibitem[Jeffreys, 1961]{jeffreys1961}
Jeffreys, H. (1961).
\newblock {\em Theory of Probability}.
\newblock Oxford University Press, Oxford, 3rd edition.

\bibitem[Jennison and Turnbull, 2000]{jennison2000}
Jennison, C. and Turnbull, B.~W. (2000).
\newblock {\em Group Sequential Methods with Applications to Clinical Trials}.
\newblock Chapman and Hall/{CRC}, Boca Raton, FL.

\bibitem[Johari et~al., 2022]{johari2022}
Johari, R., Koomen, P., Pekelis, L., and Walsh, D. (2022).
\newblock Always valid inference: Continuous monitoring of {A/B} tests.
\newblock {\em Operations Research}, 70(3):1806--1821.

\bibitem[Karalis and Macheras, 2012]{karalis2012}
Karalis, V. and Macheras, P. (2012).
\newblock Current regulatory approaches of bioequivalence testing.
\newblock {\em Expert opinion on drug metabolism \& toxicology}, 8(8):929--942.

\bibitem[Kass and Raftery, 1995]{kassraftery1995}
Kass, R.~E. and Raftery, A.~E. (1995).
\newblock Bayes factors.
\newblock {\em Journal of the American Statistical Association}, 90(430):773--795.

\bibitem[Kruschke, 2011]{kruschke2011}
Kruschke, J.~K. (2011).
\newblock Bayesian assessment of null values via parameter estimation and model comparison.
\newblock {\em Perspectives on Psychological Science}, 6(3):299--312.

\bibitem[Kruschke, 2013]{kruschke2013}
Kruschke, J.~K. (2013).
\newblock Bayesian estimation supersedes the $t$ test.
\newblock {\em Journal of Experimental Psychology: General}, 142(2):573--603.

\bibitem[Kruschke, 2015]{kruschke2015doing}
Kruschke, J.~K. (2015).
\newblock {\em Doing Bayesian Data Analysis}.
\newblock Academic Press, Boston.

\bibitem[Kruschke, 2018]{kruschke2018}
Kruschke, J.~K. (2018).
\newblock Rejecting or accepting parameter values in bayesian estimation.
\newblock {\em Advances in Methods and Practices in Psychological Science}, 1(2):270--280.

\bibitem[Lakens et~al., 2018]{lakens2018}
Lakens, D., Scheel, A.~M., and Isager, P.~M. (2018).
\newblock Equivalence testing for psychological research: A tutorial.
\newblock {\em Advances in Methods and Practices in Psychological Science}, 1(2):259--269.

\bibitem[Maxwell et~al., 2008]{maxwell2008}
Maxwell, S.~E., Kelley, K., and Rausch, J.~R. (2008).
\newblock Sample size planning for statistical power and accuracy in parameter estimation.
\newblock {\em Annual Review of Psychology}, 59.

\bibitem[McElreath, 2016]{mcelreath2016}
McElreath, R. (2016).
\newblock {\em Statistical Rethinking - A Bayesian Course with Examples in R and Stan}.
\newblock Chapman and Hall/CRC, New York.

\bibitem[Neyman and Pearson, 1933]{neymanpearson1933}
Neyman, J. and Pearson, E.~S. (1933).
\newblock On the problem of the most efficient tests of statistical hypotheses.
\newblock {\em Philosophical Transactions of the Royal Society of London. Series~A, Containing Papers of a Mathematical or Physical Character}, 231:289--337.

\bibitem[Nosek et~al., 2018]{nosek2018}
Nosek, B.~A., Ebersole, C.~R., DeHaven, A.~C., and Mellor, D.~T. (2018).
\newblock The preregistration revolution.
\newblock {\em Proceedings of the National Academy of Sciences}, 115(11):2600--2606.

\bibitem[{Open Science Collaboration}, 2015]{osc2015}
{Open Science Collaboration} (2015).
\newblock Estimating the reproducibility of psychological science.
\newblock {\em Science}, 349(6251):aac4716.

\bibitem[Rouder, 2014]{rouder2014}
Rouder, J.~N. (2014).
\newblock Optional stopping: No problem for {Bayesians}.
\newblock {\em Psychonomic Bulletin \& Review}, 21(2):301--308.

\bibitem[Sch{\"o}nbrodt and Wagenmakers, 2018]{schonbrodt2018}
Sch{\"o}nbrodt, F.~D. and Wagenmakers, E.-J. (2018).
\newblock Bayes factor design analysis: Planning for compelling evidence.
\newblock {\em Psychonomic Bulletin \& Review}, 25(1):128--142.

\bibitem[Schuirmann, 1987]{schuirmann1987}
Schuirmann, D.~J. (1987).
\newblock A comparison of the two one-sided tests procedure and the power approach for assessing the equivalence of average bioavailability.
\newblock {\em Journal of Pharmacokinetics and Biopharmaceutics}, 15(6):657--680.

\bibitem[Simmons et~al., 2011]{simmons2011}
Simmons, J.~P., Nelson, L.~D., and Simonsohn, U. (2011).
\newblock False-positive psychology: Undisclosed flexibility in data collection and analysis allows presenting anything as significant.
\newblock {\em Psychological Science}, 22(11):1359--1366.

\bibitem[{U.S. Food and Drug Administration}, 2010]{fda2010bayesian}
{U.S. Food and Drug Administration} (2010).
\newblock Guidance for the use of {Bayesian} statistics in medical device clinical trials.
\newblock Technical report, Center for Devices and Radiological Health, U.S. FDA.

\bibitem[van~der Vaart, 1998]{vandervaart1998}
van~der Vaart, A.~W. (1998).
\newblock {\em Asymptotic Statistics}.
\newblock Cambridge University Press, Cambridge.

\bibitem[Wald, 1947]{wald1947}
Wald, A. (1947).
\newblock {\em Sequential Analysis}.
\newblock John Wiley {\&} Sons, New York.

\bibitem[Wasserstein and Lazar, 2016]{wasserstein2016}
Wasserstein, R.~L. and Lazar, N.~A. (2016).
\newblock The {ASA} statement on $p$-values: Context, process, and purpose.
\newblock {\em The American Statistician}, 70(2):129--133.

\end{thebibliography}

\end{document}